\documentclass[aip, pof, preprint]{revtex4-1}
\usepackage{graphicx}
\usepackage{color}
\usepackage{amssymb}
\usepackage{amsmath}
\usepackage{tabularx}
\usepackage{bm}
\usepackage{hyperref}
\usepackage{ltxtable}
\usepackage{natbib} 
\usepackage{longtable}
\usepackage[usenames,dvipsnames]{xcolor}
\hypersetup{
colorlinks,
citecolor=blue,
filecolor=blue,
linkcolor=blue,
urlcolor=blue}

\begin{document}
\title{Scaling of large-scale quantities in Rayleigh-B\'{e}nard convection}

\author{Ambrish Pandey}
\email{ambrishiitk@gmail.com}
\author{Mahendra K. Verma}
\email{mkv@iitk.ac.in}
\affiliation{Department of Physics, Indian Institute of Technology, Kanpur 208016, India}
\date{\today}

\begin{abstract}

We derive a formula for the P\'{e}clet number ($\mathrm{Pe}$) by estimating the relative strengths of various terms of the momentum equation. Using direct numerical simulations in three dimensions we show that in the turbulent regime, the fluid acceleration is dominated by the pressure gradient, with relatively small contributions arising from the buoyancy and the viscous term; in the viscous regime, acceleration is very small due to a balance between the buoyancy and the viscous term.  Our formula for $\mathrm{Pe}$ describes the past experiments and numerical data quite well.  We also show that the ratio of the nonlinear term and the viscous term is $\mathrm{Re} \mathrm{Ra}^{-0.14}$, where $\mathrm{Re}$ and $\mathrm{Ra}$ are Reynolds and Rayleigh numbers respectively; and that the viscous dissipation rate $\epsilon_u = (U^3/d) \mathrm{Ra}^{-0.21}$, where $U$ is the root mean square velocity and $d$ is the distance between the two horizontal plates.  The aforementioned decrease in nonlinearity compared to free turbulence arises due to the wall effects.
\end{abstract}

\maketitle
\section{Introduction}\label{sec:intro}

Study of thermal convection is fundamental for the understanding of heat transport in many natural phenomena, e.g., in stars, Earth's mantle, atmospheric circulation, etc. Many researchers study Rayleigh-B\'{e}nard convection (RBC), a simplified model of convection, in which a fluid kept between two horizontal plates at a distance $d$ is heated from bottom and cooled from top~\cite{Ahlers:RMP2009, Chilla:EPJE2012, Siggia:ARFM1994, Xia:TAML2013, Lohse:ARFM2010, Bhattacharjee:Book}. Properties of the convective flow are primarily governed by two nondimensional parameters: the Prandtl number ($\mathrm{Pr}$), a ratio of the kinematic viscosity $\nu$ and the thermal diffusivity $\kappa$, and the Rayleigh number ($\mathrm{Ra}$), a ratio of the buoyancy and the viscous forces. Two important global quantities of RBC are the large-scale velocity $U$ or a dimensionless P\'{e}clet number $\mathrm{Pe} = Ud/\kappa$, and the Nusselt number $\mathrm{Nu}$, which is a ratio of the total and conductive heat transport; their dependence on $\mathrm{Ra}$ and $\mathrm{Pr}$ has been studied extensively~\cite{Ahlers:RMP2009, Chilla:EPJE2012, Siggia:ARFM1994, Xia:TAML2013}. In this paper, we derive an analytical formula for the P\'{e}clet number  that can explain the experimental and numerical results quite well.  The formula  however involves certain coefficients that are determined using numerical simulations.  In addition to $\mathrm{Pe}$, we also discuss the scaling of Nusselt number and dissipation rates.
 
Many researchers\cite{Malkus:PRSL1954a,Malkus:PRSL1954b, Kraichnan:POF1962, Castaing:JFM1989, Shraiman:PRA1990, Siggia:ARFM1994, Cioni:JFM1997, Grossmann:JFM2000, Grossmann:PRL2001, Grossmann:PRE2002, Grossmann:POF2004, Grossmann:POF2011, Stevens:JFM2013} have studied  the Nusselt and Reynolds numbers. Using the arguments of marginal stability theory,  Malkus~\cite{Malkus:PRSL1954a, Malkus:PRSL1954b} deduced that $\mathrm{Nu} \approx (\mathrm{Ra/Ra}_c)^{1/3}$ by assuming that the heat transport is independent of $d$. Using mixing length theory, Kraichnan~\cite{Kraichnan:POF1962} proposed that for very large Rayleigh numbers,  the heat transport is independent of kinematic viscosity and thermal diffusivity of the fluid. The boundary layers in this ``ultimate regime"  becomes turbulent leading to $\mathrm{Nu \sim \sqrt{RaPr}}$ and $\mathrm{Re \sim \sqrt{Ra/Pr}}$.

Castaing \textit{et al.}~\cite{Castaing:JFM1989} performed experiments with helium gas ($\mathrm{Pr} \approx 0.7$) and observed $\mathrm{Nu \sim Ra^{0.28}}$, and a Reynolds number $\mathrm{Re =Pe/Pr} \sim \mathrm{Ra}^{0.49}$ based on the peak frequency of the power spectrum. Sano \textit{et al.}~\cite{Sano:PRA1989} measured a P\'{e}clet number based on the mean vertical velocity near the side-wall and found that $\mathrm{Pe} \sim \mathrm{Ra}^{0.48}$. Castaing \textit{et al.}~\cite{Castaing:JFM1989} proposed existence of a mixing zone where hot rising plumes meet mildly warm fluid. By matching the velocity of the hot fluid at the end of the mixing zone with those of the central region, Castaing \textit{et al.}~\cite{Castaing:JFM1989} argued that $\mathrm{Nu \sim Ra^{2/7}}$, $\mathrm{Re}_c \sim \mathrm{Ra}^{3/7}$, where $\mathrm{Re}_c$ is based on the typical velocity scale in the central region. Using the properties of the boundary layer, Shraiman and Siggia~\cite{Shraiman:PRA1990}  derived that $\mathrm{Nu \sim Pr^{-1/7} Ra^{2/7}}$ and $\mathrm{Re \sim Pr^{-5/7} Ra^{3/7}}[2.5 \ln(\mathrm{Re}) + 5]$. They also derived exact relations between the Nusselt number and the global viscous ($\epsilon_u$) and thermal ($\epsilon_T$) dissipation rates~\cite{Siggia:ARFM1994}.

One of the most recent and popular models of large-scale quantities of RBC is by Grossmann and Lohse~\cite{Grossmann:JFM2000, Grossmann:PRL2001, Grossmann:PRE2002, Grossmann:POF2004, Grossmann:POF2011} (henceforth referred to as GL theory).  In the Shraiman and Siggia's~\cite{Shraiman:PRA1990} exact relations connecting the dissipation rates with the Nusselt  and Reynolds numbers, Grossmann and Lohse~\cite{Grossmann:JFM2000, Grossmann:PRL2001} substituted the contributions from the bulk and the boundary layers.  This process enabled  Grossmann and Lohse to derive different formulae for the Nusselt and Reynolds numbers in the bulk and boundary-layer dominated regimes.  The coefficients of the formulae were determined using experimental and simulation inputs.    Later Stevens \textit{et al.}~\cite{Stevens:JFM2013} updated the  coefficients  by including more recent simulation and experimental data. GL theory has been quite successful in explaining the heat transport and Reynolds number in many numerical simulations and experiments.  In this paper we derive a formula for the P\'{e}clet number using a different approach; we will contrast the differences between our model and GL towards the end of the paper.

The Reynolds number has been measured in many experiments and direct numerical simulations (DNS) for a vast range of Rayleigh and Prandtl numbers, and it can be quantified in various ways: based on the maximum velocity of the horizontal velocity profiles~\cite{Xin:PRE1997, Lam:PRE2002}, absolute peak value of the vertical velocity~\cite{Silano:JFM2010, Horn:JFM2013}, the root mean square (rms) velocity~\cite{Lam:PRE2002, Verma:PRE2012, Scheel:JFM2014, Pandey:PRE2014}, etc. It can also be computed using the peak frequency in power spectra of the temperature or velocity cross-correlation functions~\cite{Cioni:JFM1997,Qiu:PRL2001, Lam:PRE2002, Niemela:JFM2001}.  Based on these estimates, Cioni \textit{et al.}~\cite{Cioni:JFM1997} reported that $\mathrm{Re \sim Ra^{0.42}}$ for mercury ($\mathrm{Pr} \approx 0.022$), and Qiu and Tong~\cite{Qiu:PRL2001} reported that $\mathrm{Re \sim Ra^{0.46}}$ for water ($\mathrm{Pr} \approx 5.4$). Lam \textit{et al.}~\cite{Lam:PRE2002} studied the Nusselt and Reynolds number scaling using experiments with organic fluids and measured $\mathrm{Re}$ based on the oscillation frequency in large-scale flow.  They showed that $\mathrm{Re} \sim \mathrm{Ra^{0.43} Pr^{-0.76}}$ for $3 \leq \mathrm{Pr} \leq 1205$ and $10^8 \leq \mathrm{Ra} \leq 3 \times 10^{10}$. Based on the volume-averaged rms velocity in numerical simulations, Verma \textit{et al.}~\cite{Verma:PRE2012} observed that $\mathrm{Pe}$ scales as $\mathrm{Ra}^{0.43}$ and $\mathrm{Ra}^{0.49}$ respectively for $\mathrm{Pr} = 0.2$ and $6.8$, and Scheel and Schumacher~\cite{Scheel:JFM2014} found $\mathrm{Re \sim Ra^{0.49}}$ for $\mathrm{Pr} = 0.7$. In DNS of very large Prandtl numbers, Silano \textit{et al.}~\cite{Silano:JFM2010}, Horn \textit{et al.}~\cite{Horn:JFM2013} and Pandey \textit{et al.}~\cite{Pandey:PRE2014} observed that $\mathrm{Re} \sim \mathrm{Ra}^{0.60}$.

In many experimental and numerical investigations~\cite{Ahlers:RMP2009, Siggia:ARFM1994, Chilla:EPJE2012, Xia:TAML2013, Kraichnan:POF1962, Castaing:JFM1989, Kerr:JFM1996, Cioni:JFM1997, Chavanne:PRL1997, Camussi:POF1998, Verzicco:JFM1999, Glazier:Nature1999, Niemela:Nature2000, Grossmann:JFM2000, Grossmann:PRL2001, Grossmann:PRE2002, Grossmann:POF2004, Grossmann:POF2011, Stevens:JFM2013, Lohse:PRL2003, Roche:PRE2001, Niemela:JFM2003, Xia:PRL2002, Shishkina:JFM2009, Stevens:JFM2010, Stevens:JFM2011, Silano:JFM2010, Verma:PRE2012, Horn:JFM2013, Scheel:JFM2012, Scheel:JFM2014, Pandey:PRE2014, Ahlers:NJP2009, Ahlers:NJP2012},  the Nusselt number scales as $\mathrm{Nu \sim Ra^{\gamma}}$, where $\gamma$ has been observed from 0.25 to 0.50.  The exponent of 0.50 has been reported for numerical experiments with periodic boundary condition,~\cite{Lohse:PRL2003, Verma:PRE2012} and in turbulent free convection due to density gradient~\cite{Cholemari:JFM2009}.  A possible transition to the ultimate regime has been reported in some experiments~\cite{Chavanne:PRL1997, Roche:PRE2001, Ahlers:NJP2009, Ahlers:NJP2012, He:NJP2012, He:JFM2016}, while some others did not find any signature of a  transition to the ultimate regime~\cite{Glazier:Nature1999, Niemela:Nature2000, Urban:PRL2011, Urban:PRL2012, Skrbek:JFM2015}. The Prandtl number dependence of the heat transport has also been investigated in simulations~\cite{Verzicco:JFM1999, Stevens:JFM2011} and experiments~\cite{Ashkenazi:PRL1999, Xia:PRL2002}. Verzicco and Camussi~\cite{Verzicco:JFM1999} found $\mathrm{Nu \sim Pr^{0.14}}$ for $\mathrm{Pr} \leq 0.35$ and no variation beyond $\mathrm{Pr} = 0.35$. Xia \textit{et al.}~\cite{Xia:PRL2002} observed that the heat transport decreases weakly  with the increase of $\mathrm{Pr}$ yielding  $\mathrm{Nu \sim Ra^{0.30} Pr^{-0.03}}$ for $4 \leq \mathrm{Pr} \leq 1353$. 

In RBC, the thermal plates induce anisotropy and sharp gradients in the flow. For example, the maximum drop in the temperature  occurs mostly near the top and bottom plates, whereas the temperature remains an approximate constant in the central region~\cite{Shishkina:JFM2009}. Similarly, Emran and Schumacher~\cite{Emran:JFM2008, Emran:EPJE2012} and Stevens \textit{et al.}~\cite{Stevens:JFM2010} reported that the thermal and the viscous dissipation rates in the boundary layers exceeds those in the bulk. In this paper we compute the volume-averaged viscous and thermal dissipation rates, and show that RBC has a  lower nonlinearity compared to homogeneous and isotropic flows of free or unbound turbulence.

In this paper we quantify various terms in the momentum  equation and obtain an analytical relation for $\mathrm{Pe(Ra,Pr)}$.  The formula depends on certain coefficients that are determined using numerical simulations.  Our derivation of $\mathrm{Pe}$, which is very different from that of Grossman and Lohse\cite{Grossmann:JFM2000, Grossmann:PRL2001}, has a single formula for $\mathrm{Pe}$. We show in this paper that the predictions of our formula match with  most of the experimental and numerical simulations.  In this paper we also discuss the $\mathrm{Pr}$ and $\mathrm{Ra}$ dependence of the Nusselt number and the dissipation rates in RBC.  Our analysis also shows that in the turbulent regime, the acceleration of a fluid parcel is dominated by the pressure gradient.  However in the viscous regime, the most dominant terms, the buoyancy and the viscous force balance each other.

The outline of the paper is following. Section~\ref{sec:equations} contains the details about the governing equations. In Sec.~\ref{sec:temp}, we discuss the properties of the average temperature profile in RBC. In Sec.~\ref{sec:Pe}, we construct a model to compute $\mathrm{Pe}$ as a function of $\mathrm{Ra}$ and $\mathrm{Pr}$. Simulation details and comparison of our model predictions with earlier results are discussed in Sec.~\ref{sec:no_slip}, and the scaling of Nusselt number and normalized thermal and viscous dissipation rates are presented in Sec.~\ref{sec:nu_diss}. Section~\ref{sec:free_slip} contains the results of RBC simulations with free-slip boundary condition.  We conclude in Sec.~\ref{sec:concl}.

\section{Governing equations}
\label{sec:equations}
The equations of Rayleigh-B\'{e}nard convection under the Boussinesq approximation for a fluid confined between two plates separated by a distance $d$ are 
\begin{eqnarray}
\frac{\partial {\bf u}}{\partial t} + ({\bf u} \cdot \nabla) {\bf u} & = & -\frac{\nabla \sigma}{\rho_0} + \alpha g \theta \hat{\mathbf{z}} + \nu \nabla^2 {\bf u},  \label{eq:u} \\
\frac{\partial  \theta}{\partial t} + ({\bf u} \cdot \nabla) \theta & = & \frac{\Delta}{d} u_z + \kappa \nabla^2 \theta, \label{eq:th} \\
\nabla \cdot {\bf u} & = & 0, \label{eq:cont}
\end{eqnarray}
where ${\bf u} = (u_x, u_y, u_z)$ is the velocity field, $\theta$ and $\sigma$ are the deviations of temperature and pressure from the conduction state, $\rho_0, \alpha, \kappa$, and $\nu$ are respectively the mean density, the heat expansion coefficient, the thermal diffusivity and the kinematic viscosity of the fluid, $\Delta$ is the temperature difference between top and bottom plates, $g$ is the gravitational acceleration, and $\hat{\bf z}$ is the unit vector in the upward direction. 

The two nondimensional parameters of RBC are the Rayleigh number $\mathrm{Ra} = \alpha g \Delta d^3 /\nu \kappa$ and the Prandtl number $\mathrm{Pr} = \nu/\kappa$. A nondimensionalized version of the above equations using $d$ as the length scale, $\sqrt{\alpha g \Delta d}$ as the velocity scale, $\Delta$ as the temperature scale, and $d/\sqrt{\alpha g \Delta d}$ as the time scale is
\begin{eqnarray}
\frac{\partial \bf u'}{\partial t'} + (\bf u' \cdot \nabla') \bf u' & = & -\nabla' \sigma' + \theta' \hat{\bf z} + \sqrt{\frac{\mathrm{Pr}}{\mathrm{Ra}}} \nabla'^2 \bf u', \label{eq:u_ndim} \\
\frac{\partial \theta'}{\partial t'} + (\bf u' \cdot \nabla') \theta' & = &  u'_z + \frac{1}{\sqrt{\mathrm{Ra}\mathrm{Pr}}}\nabla'^2 \theta', \label{eq:th_ndim} \\
\nabla' \cdot \bf u' & = & 0. \label{eq:inc_ndim}
\end{eqnarray}
Here the primed variables represent dimensionless quantities. The magnitude of the large-scale velocity is computed using the time-averaged total kinetic energy $E_u$ as $U = \sqrt{2 \langle E_u \rangle_t}$, where $\langle \rangle_t$ denotes the averaging over time. The P\'{e}clet number is the ratio of the advection term and the diffusion term of the temperature equation, and it is defined as 
\begin{equation}
\mathrm{Pe} = \frac{|{\bf u} \cdot \nabla \theta|}{|\kappa \nabla^2 \theta|} =  \frac{U d}{\kappa}.
\end{equation}
P\'{e}clet number is analogous to Reynolds number, which is the ratio of the nonlinear term and the viscous term of the momentum equation.

In this paper, we study the rms values of the large-scale velocity and temperature fields, and other related global quantities like the Nusselt number and the dissipation rates.

\section{Temperature profile and boundary layer}
\label{sec:temp}
\begin{figure}
\includegraphics[scale=1.1]{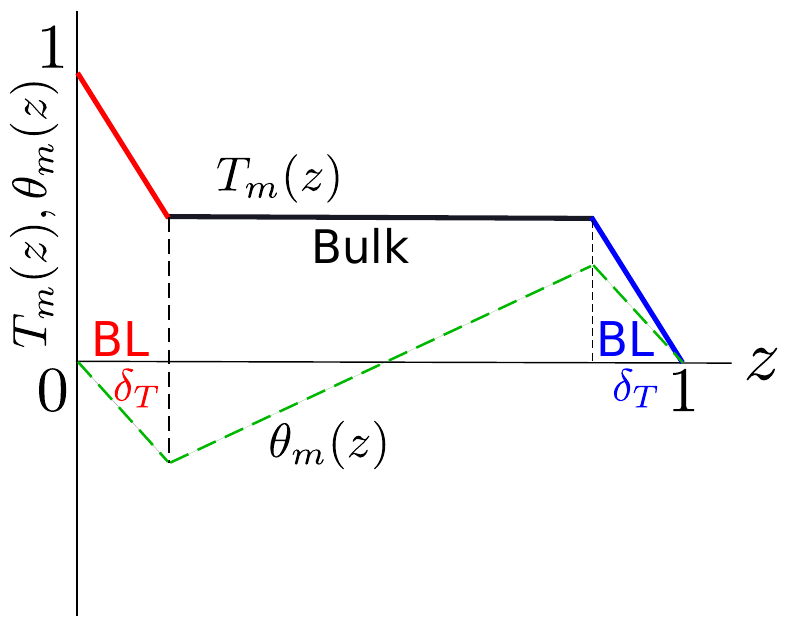}
\caption{A schematic diagram of the planar-averaged temperature  as a function of the vertical coordinate. The temperature drops sharply to $1/2$ in the boundary layers.}
\label{fig:Tprofile}
\end{figure}

The temperature $T(x,y,z)$ in a Rayleigh-B\'{e}nard cell fluctuates in time, and  it can be decomposed into a conductive profile and  fluctuations superimposed on it, i.e.,
 \begin{equation}
T(x,y,z)  = T_c(z) + \theta(x,y,z) =  1- z+ \theta(x,y,z).
\label{eq:theta}
\end{equation}
Here we work with a nondimensionalized system for which the bottom and the top plates are separated by a unit distance, and are kept at temperatures 1 and 0 respectively.  We define the planar average of temperature, $T_m(z) = \langle T \rangle_{xy}$.  Experiments and numerical simulations reveal that $T_m(z) \approx 1/2$ in the bulk, and it drops abruptly in the boundary layers near the top and bottom plates~\cite{Emran:JFM2008, Shishkina:JFM2009}, as shown in Fig.~\ref{fig:Tprofile}.  The quantitative expression for $T_m(z)$ can be approximated as
\begin{equation}
    T_m(z)= 
\begin{cases}
    1 - \frac{z}{2\delta_T}, & \text{if } 0 < z < \delta_T \\
	1/2 , & \text{if } \delta_T < z < 1 - \delta_T \\
	\frac{1-z}{2\delta_T} , & \text{if } 1 - \delta_T < z < 1 
\end{cases}
\end{equation}
where $\delta_T$ is the thickness of the thermal boundary layer.  

Horizontal averaging of Eq.~(\ref{eq:theta}) yields
 \begin{equation}
\theta_m(z) = T_m(z) + z-1,
\label{eq:theta(z)}
\end{equation}
where $\theta_m(z)$ is
\begin{equation}
    \theta_m(z) = 
\begin{cases}
    z \left( 1 - \frac{1}{2\delta_T} \right ), & \text{if } 0 < z < \delta_T \\
	z - 1/2, & \text{if } \delta_T < z < 1 - \delta_T \\
    (z-1)\left( 1 - \frac{1}{2\delta_T} \right ), & \text{if } 1 - \delta_T < z < 1 
\end{cases}
\end{equation}
as exhibited in Fig.~\ref{fig:Tprofile}.  Now we compute the Fourier transform of  $\theta_m(z)$.  For thin boundary layers, the Fourier transform $\hat{\theta}_m(0,0,k_z)$ is dominated by the contributions from the bulk that yields
\begin{eqnarray}
\hat{\theta}_m(0,0,k_z) & = & \int_0^1 \theta_m(z) \sin(k_z \pi z) dz \nonumber \\
 & \approx & \int_0^1 (z-1/2) \sin(k_z \pi z) dz  \nonumber\\
 & \approx & \begin{cases}
       -\frac{1}{\pi k_z} &\mathrm{for \, even} \, k_z \\
        0 \, &\mathrm{otherwise}
\end{cases}
\label{eq:theta_hat_k}
\end{eqnarray}
The above result plays a crucial role in the scaling of global quantities, as we will show below.  Earlier, Mishra and Verma~\cite{Mishra:PRE2010} and Pandey \textit{et al.}~\cite{Pandey:PRE2014} had observed the above features in numerical simulations; Mishra and Verma~\cite{Mishra:PRE2010} had explained it using energy transfer arguments on the Fourier modes $\hat{\theta}(2n,0, 2n)$.  A consequence of Eq.~(\ref{eq:theta_hat_k}) is that the entropy spectrum 
\begin{equation}
E_\theta(k) = \sum_{k \le k' < k+1} \frac{1}{2} |\theta({\bf k'})|^2
\end{equation}
exhibits a dual branch---$k^{-2}$ corresponding to Eq.~(\ref{eq:theta_hat_k}) as the first branch, and a second branch for the rest of the $\hat{\theta}$ modes~\cite{Mishra:PRE2010, Pandey:PRE2014}.

It is interesting to note that the corresponding velocity mode,  $\hat{u}_z(0, 0, k_z) = 0$ because of the incompressibility condition ${\bf k} \cdot \hat{\bf u}(0, 0, k_z) = k_z \hat{u}_z(0,0,k_z) = 0$.  Also, $\hat{u}_x(0, 0, k_z) = \hat{u}_y(0, 0, k_z) =  0$ in the absence of a  mean flow along the horizontal direction.  Hence for ${\bf k} = (0,0,k_z)$ modes, the momentum equation yields
\begin{equation}
0  = -\frac{ i {\bf k} \hat{\sigma}({\bf k})}{\rho_0} + \alpha g \hat{\theta}({\bf k}) \hat{\bf z} 
\end{equation}
or $d \sigma_m(z) /dz = \rho_0 \alpha g \theta_m$.   The dynamics of the remaining set of Fourier modes is governed by the momentum equation as
\begin{equation}
\frac{\partial \hat{{\bf u}}({\bf k})}{\partial t} + i \sum_{\bf{p+q=k}} {[\bf k \cdot \hat{u}}({\bf q})] {\bf \hat{u}}({\bf p})  = -\frac{  i {\bf k} \hat{\sigma}_{\mathrm{res}}({\bf k})}{\rho_0} + \alpha g \hat{\theta}_\mathrm{res}({\bf k}) \hat{\bf z} - \nu k^2 \hat{{\bf u}}({\bf k}), \label{eq:u_four}
\end{equation}
where
\begin{eqnarray}
\theta & = & \theta_\mathrm{res} + \theta_m, \label{eq:theta_res} \\
\sigma & = & \sigma_\mathrm{res} + \sigma_m. \label{eq:sigma_res} 
\end{eqnarray}
Hence,  the modes $\hat{\theta}_m(0,0, k_z)$ and $\hat{\sigma}_m(0,0, k_z)$ do not couple with the velocity modes in the momentum equation, but $\theta_\mathrm{res}$ and $\sigma_\mathrm{res}$ do.

In the next section, we will quantify the large-scale velocity in RBC.

\section{Universal formula for $U$ or P\'{e}clet number} 
\label{sec:Pe}

We derive an expression for the large-scale velocity $U$ from the momentum equation [Eq.~(\ref{eq:u})].  According to this equation,  the material acceleration $D{\bf u}/Dt$ of a fluid element results from the pressure gradient, buoyancy, and the viscous force.  Under steady state, we assume that $\langle \partial {\bf u}/\partial t \rangle \approx 0$, hence,  a dimensional analysis of the momentum equation yields
\begin{equation}
c_1 \frac{U^2}{d} =  c_2  \frac{U^2}{d}+  c_3 \alpha g \theta_{\mathrm{res}} - c_4 \nu \frac{U}{d^2}, \label{eq:U}
\end{equation}
where  $c_i$'s are dimensionless coefficients. We  observe in our numerical simulations (to be discussed later) that the pressure gradient provides the acceleration to a fluid parcel whereas the viscous force opposes the motion. Therefore we  choose the sign of $c_2$ same as that of $c_1$, and the sign of $c_4$ has been chosen opposite to those of $c_1$ and $c_2$.  In RBC, buoyancy provides additional acceleration, hence $c_3$ has the same sign as $c_1$ and $c_2$.

As discussed in the previous section,  the momentum equation contains $\theta_\mathrm{res} =\theta-\theta_m$, not $\theta$.    The coefficients are defined as
\begin{eqnarray}
c_1 & = & \frac{|{\bf u \cdot \nabla u}|} {U^2/d}, \nonumber \\
c_2 & = & \frac{ |\nabla \sigma|_{\mathrm{res}} /\rho_0} {U^2/d}, \nonumber \\
c_3 & = & |\theta_{\mathrm{res}}/\Delta|, \nonumber \\
c_4 & = & \frac{|\nabla^2 {\bf u}|} {U/d^2}.
\end{eqnarray}
We will show later that $c_i$'s are functions of $\mathrm{Ra}$ and $\mathrm{Pr}$ that yields very interesting and nontrivial scaling relations.  Note that typical dimensional arguments in fluid mechanics assume $c_i$'s to be constants, which is valid for free or unbounded turbulence.  

Multiplication of Eq.~(\ref{eq:U}) with $d^3/\kappa^2$ yields
\begin{equation}
c_1 \mathrm{Pe}^2 = c_2 \mathrm{Pe}^2 + c_3 \mathrm{RaPr} - c_4 \mathrm{PePr},
\label{eq:Pe_eqn}
\end{equation}
whose solution is
\begin{equation}
\mathrm{Pe} = \frac{-c_4 \mathrm{Pr} + \sqrt{c_4^2 \mathrm{Pr}^2 + 4(c_1-c_2) c_3\mathrm{RaPr}}}{2 (c_1-c_2)}. \label{eq:Pe_analy}
\end{equation}	
Now we can compute the P\'{e}clet number using Eq.~(\ref{eq:Pe_analy}) given $c_i(\mathrm{Pr}, \mathrm{Ra})$.  We compute these coefficients in subsequent sections. We remark that $\mathrm{Pe}$ could be a function of geometrical factors and aspect ratio.  

In the viscous regime, the nonlinear term, ${\bf u \cdot \nabla u}$, and the pressure gradient,  $-\nabla \sigma$, are much smaller than the buoyancy and the viscous terms, hence in this regime
\begin{equation}
 c_3 \mathrm{RaPr} - c_4 \mathrm{PePr} \approx 0,
\end{equation}	
which yields
\begin{equation}
\mathrm{Pe} \approx \frac{c_3}{c_4} \mathrm{Ra}.
\label{eq:Pe_visc}
\end{equation}	
We can deduce the properties under the turbulent regime by ignoring the viscous term in Eq.~(\ref{eq:Pe_eqn}), which yields
\begin{equation}
c_1 \mathrm{Pe}^2 \approx c_2 \mathrm{Pe}^2 + c_3 \mathrm{RaPr}.
\label{eq:Pe_turb_cond}
\end{equation}
The solution of the above equation is
\begin{equation}
\mathrm{Pe} \approx \sqrt{\frac{c_3}{|c_1-c_2|} \mathrm{RaPr}}. 
\label{eq:Pe_turb}
\end{equation}	
The above two limiting expressions of $\mathrm{Pe}$ can be derived from Eq.~(\ref{eq:Pe_analy}).  We obtain turbulent regime when  
 \begin{equation}
\mathrm{Ra} \gg \frac{c_4^2 \mathrm{Pr}}{4 c_3 |c_1-c_2|}  
\label{eq:Ra_cond_turb}
\end{equation}	
and viscous regime for $\mathrm{Ra} \ll  c_4^2 \mathrm{Pr}/(4 c_3 |c_1-c_2|)$.  We will examine these cases once we deduce the forms of $c_i$'s using numerical simulations.

In the next section, we  compute the coefficients $c_i$'s   using our numerical simulation. Then we predict the functional dependence of $\mathrm{Pe}(\mathrm{Ra},\mathrm{Pr})$.

\section{Numerical simulation and results}
\label{sec:no_slip}
\begin{figure}
\includegraphics[scale=1]{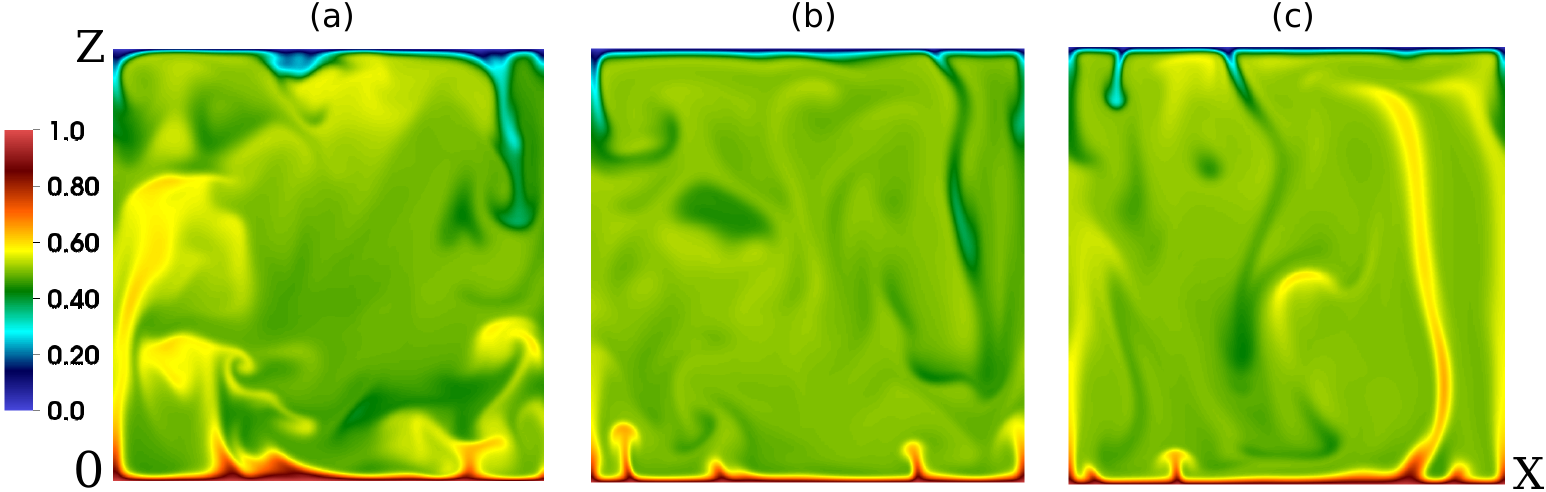}
\caption{For the no-slip boundary condition, the instantaneous temperature field in a vertical cross section at $y = 0.4$ for $\mathrm{Ra} = 5 \times 10^7$ and (a) $\mathrm{Pr} = 1$, (b) $\mathrm{Pr} = 6.8$, and (c) $\mathrm{Pr} = 10^2$.  The flow structures or plumes get sharper as $\mathrm{Pr}$ increases. }
\label{fig:temp_field_ns}
\end{figure}

We perform RBC simulations by solving Eqs.~(\ref{eq:u_ndim}--\ref{eq:inc_ndim}) in a three-dimensional unit box  for $\mathrm{Pr} = 1, 6.8$, and $10^2$ and $\mathrm{Ra}$ between $10^6$ and $5 \times 10^8$ using an open-source finite-volume code {\sc OpenFoam}~\cite{OpenFOAM}. We employ no-slip boundary condition for the velocity field at all the walls. For the temperature field, we impose isothermal condition on the top and bottom plates, and adiabatic condition at the vertical walls. The time stepping is performed using the second-order Crank-Nicolson method. Total number of grid points in our simulations vary from $60^3$ and $256^3$ with  finer grids employed near the boundary layers\cite{Grotzbach:JCP1983, Shishkina:NJP2010}.  We employ nonuniform mesh with higher concentration of grid points,  from 4 to 6, near the boundaries.  We validate our code by comparing the  Nusselt number with those computed in earlier numerical simulations and experiments.  We  show that our results are grid-independent by showing that for $\mathrm{Pr} = 1$ and $\mathrm{Ra} = 10^8$, the Nusselt numbers for $100^3$ and $256^3$ grids are close to each other within 3\%. Figure~\ref{fig:temp_field_ns} shows the temperature field in a vertical $xz$-plane at $y = 0.4$ for $\mathrm{Pr} = 1, 6.8, 10^2$, and $\mathrm{Ra} = 5 \times 10^7$. Figures~\ref{fig:temp_field_ns}(b,c) exhibit \textit{mushroom-shaped} sharper plumes, characteristics of large Prandtl number RBC.~\cite{Chilla:EPJE2012}

\begin{figure}
\includegraphics[scale=1.1]{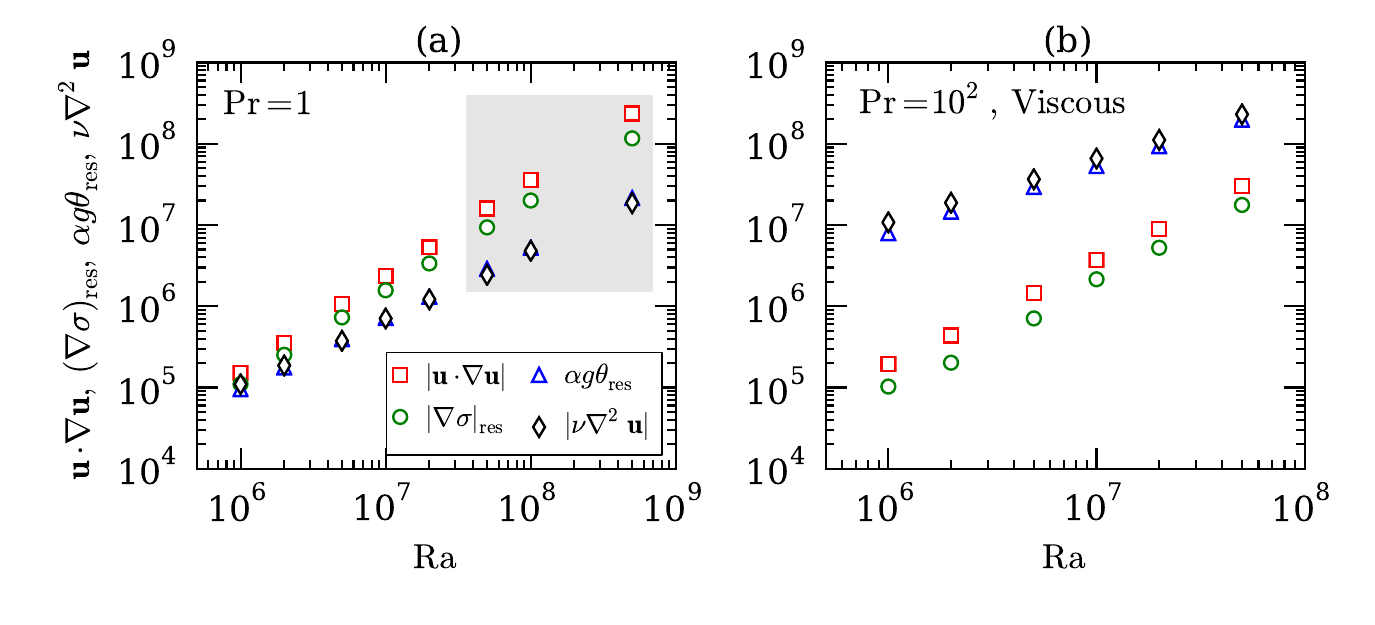}
\caption{For the no-slip boundary condition: the comparison of the rms values of ${\bf u} \cdot \nabla {\bf u}$, $(-\nabla \sigma)_{\mathrm{res}}$, $\alpha g \theta_{\mathrm{res}} \hat{\bf z}$, and $\nu \nabla^2 {\bf u}$ as function of $\mathrm{Ra}$ for (a) $\mathrm{Pr} = 1$ and (b) $\mathrm{Pr} = 10^2$.  The shaded region of Fig.~(a) corresponds to the turbulent regime, while all the runs of $\mathrm{Pr}=10^2$ belong to the viscous regime.}
\label{fig:compare_ns}
\end{figure}

Table~\ref{table:details_ns} summarizes our simulation parameters, as well as the P\'{e}clet and Nusselt numbers, and $k_{\mathrm{max}} \eta_\theta$.  For most of our runs, $k_{\mathrm{max}} \eta_\theta \ge 1$, where $\eta_\theta$ is the Batchelor scale. The Batchelor scale $\eta_\theta = (\kappa^3/\epsilon_u)^{1/4}$ is related to the Kolmogorov scale $\eta_u$ as $\eta_\theta = \eta_u \mathrm{Pr}^{-3/4}$. For $\mathrm{Pr} \ge 1$, $\eta_\theta \le \eta_u$, and therefore the mean grid spacing should be smaller than $\eta_\theta$.  We continue  the simulation till it reaches statistical steady state, and then we compute averages of the rms values of $|{\bf u} \cdot \nabla {\bf u}|$, $|(-\nabla \sigma)_{\mathrm{res}}|$, $|\alpha g \theta_{\mathrm{res}} \hat{\bf z}|$ and $|\nu \nabla^2 {\bf u}|$. We compute these quantities by first taking a volume average over the entire box and then taking a time average. We perform these computations for a wide range of $\mathrm{Pr}$ and  $\mathrm{Ra}$ and plot them as function of $\mathrm{Ra}$ in Fig.~\ref{fig:compare_ns} for $\mathrm{Pr} = 1$ and $\mathrm{Pr} = 100$. The $\mathrm{Ra}$-dependence of $|{\bf u} \cdot \nabla {\bf u}|$, $|(-\nabla \sigma)_{\mathrm{res}}|$, $|\alpha g \theta_{\mathrm{res}} \hat{\bf z}|$ and $|\nu \nabla^2 {\bf u}|$ are listed in Table~\ref{table:terms_ns}.

\begin{table*}
\begin{ruledtabular}
\caption{Details of our simulations with no-slip boundary condition: $N^3$ is the total number of grid points. }
\begin{tabular}{cccccc|cccccc}
$\mathrm{Pr}$ & $\mathrm{Ra}$ & $N^3$ & Nu & Pe & $k_{\mathrm{max}}\eta_\theta$ & $\mathrm{Pr}$ & $\mathrm{Ra}$ & $N^3$ & Nu & Pe & $k_{\mathrm{max}}\eta_\theta$ \\ \hline
1 & $1 \times 10^6$ & $60^3$ & 8.0 & 146.1  & 3.8 & 6.8 & $5 \times 10^6$ & $60^3$ & 13.1 & 413.6 &  1.4 \\
1 & $2 \times 10^6$ & $60^3$ & 10.0 & 211.3 & 3.0 & 6.8 & $1 \times 10^7$ & $80^3$ & 16.2 & 608.6 &  1.5 \\
1 & $5 \times 10^6$ & $60^3$ & 13.4 & 340.3 & 2.3 & 6.8 & $2 \times 10^7$ & $80^3$ & 20.3 & 903.2 &  1.2 \\
1 & $1 \times 10^7$ & $80^3$ & 16.3 & 485.4  & 2.4 & 6.8 & $5 \times 10^7$ & $80^3$ & 27.7 & 1536  & 0.8 \\
1 & $2 \times 10^7$ & $80^3$ & 20.2 & 687.4  & 1.9 & $10^2$ & $1 \times 10^6$ & $60^3$ & 8.5 & 190.7 & 1.2 \\
1 & $5 \times 10^7$ & $80^3$ & 26.8 & 1103 &  1.5 & $10^2$ & $2 \times 10^6$ & $60^3$ & 11.2 & 278.2 & 0.9 \\
1 & $1 \times 10^8$ & $100^3$ & 32.9 & 1554 &  1.4 & $10^2$ & $5 \times 10^6$ & $60^3$ & 14.5 & 500.0  & 0.7 \\
1 & $1 \times 10^8$ & $256^3$ & 31.9 & 1537 &  3.5 & $10^2$ & $1 \times 10^7$ & $80^3$ & 17.1 & 704.2  & 0.7 \\
1 & $5 \times 10^8$ & $256^3$ & 51.2 & 3408 &  2.1 & $10^2$ & $2 \times 10^7$ & $80^3$ & 20.7 & 1044  & 0.6 \\
6.8 & $1 \times 10^6$ & $60^3$ & 8.4 & 182.7 & 2.3 & $10^2$ & $5 \times 10^7$ & $80^3$ & 27.7 & 1826 & 0.4 \\
6.8 & $2 \times 10^6$ & $60^3$ & 9.9 & 252.8 &  1.9 & -- & -- & -- & -- & -- & --	 \\
\end{tabular}
\label{table:details_ns}
\end{ruledtabular} 
\end{table*}

\begin{figure}
\includegraphics[scale=1.2]{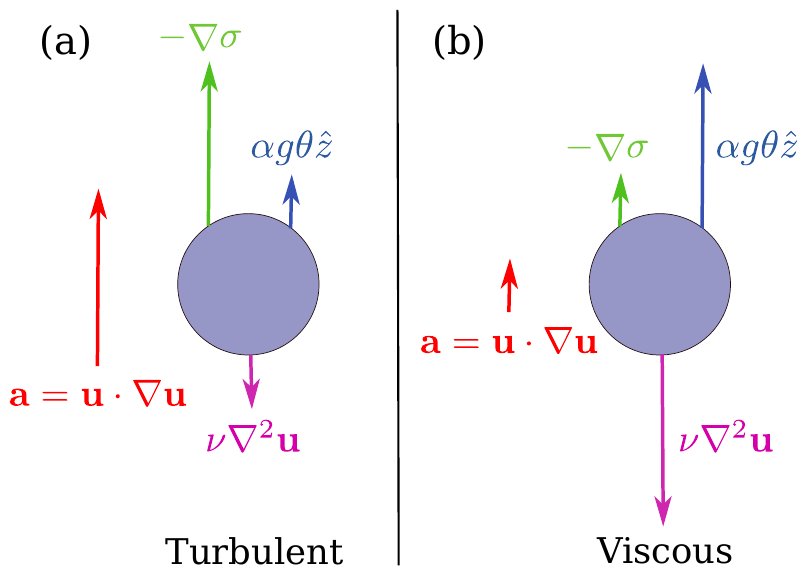}
\caption{For the no-slip boundary condition, the relative strengths of the forces acting on a fluid parcel. In the turbulent regime, the acceleration ${\bf u} \cdot \nabla {\bf u}$ is provided primarily by the pressure gradient.  In the viscous regime, the buoyancy and  the viscous force dominate the pressure gradient,  and they balance each other.}
\label{fig:schematic_ns}
\end{figure}

We observe that for $\mathrm{Pr} = 1$ and $\mathrm{Ra}$ near $10^8$ [the shaded region of Fig.~\ref{fig:compare_ns}(a)], the nonlinear term ($|{\bf u \cdot \nabla u}|$) and the pressure gradient ($|\nabla \sigma|$) are much larger than the viscous and the buoyancy terms.   It is evident from the fact that the Reynolds number for $ \mathrm{Ra} = 5 \times 10^7, 10^8, 5\times 10^8$ are  approximately 1103, 1537, and 3408 respectively.  In the other limit, for $\mathrm{Pr} = 100$ [Fig~\ref{fig:compare_ns}(b)],  the viscous force and buoyancy  are always larger than the nonlinear term and the pressure gradient.  We depict the force balance in Fig.~\ref{fig:schematic_ns}.  Our numerical results are consistent with the intuitive pictures of the turbulent and viscous flows.

\begin{table}
\caption{Scaling of various terms of the momentum equation (scaled as $\kappa^2/d^3$) for the no-slip boundary condition. The errors in the exponents are approximately $0.02$.}
\begin{ruledtabular}
\begin{tabular}{ccc}
	& Turbulent regime & Viscous regime \\
\hline 
$|{\bf u} \cdot \nabla {\bf u}|$ & $\mathrm{Ra}^{1.2}$ & $\mathrm{Ra}^{1.3}$ \\
$|(-\nabla \sigma)_{\mathrm{res}}|$ & $\mathrm{Ra}^{1.1}$ & $\mathrm{Ra}^{1.3}$ \\
$|\alpha g \theta_{\mathrm{res}}|$ & $\mathrm{Ra}^{0.87}$ & $\mathrm{Ra}^{0.82}$ \\
$|\nu \nabla^2 {\bf u}|$ & $\mathrm{Ra}^{0.87}$ & $\mathrm{Ra}^{0.82}$ \\
\end{tabular}
\end{ruledtabular}
\label{table:terms_ns}
\end{table}

\begin{table}
\setlength{\tabcolsep}{1.2cm}
\caption{Functional dependence of the coefficients $c_i$'s on $\mathrm{Ra}$ and $\mathrm{Pr}$ under the no-slip boundary condition.}
\begin{tabular}{lc|lc}
\hline \hline
$c_1$ & $1.5 \mathrm{Ra}^{0.10} \mathrm{Pr}^{-0.06}$ & $c_3$ & $0.75 \mathrm{Ra}^{-0.15} \mathrm{Pr}^{-0.05}$  \\
$c_2$ & $1.6 \mathrm{Ra}^{0.09} \mathrm{Pr}^{-0.08}$ & $c_4$ & $20 \mathrm{Ra}^{0.24} \mathrm{Pr}^{-0.08}$ \\
\hline \hline	
\end{tabular}
\label{table:c_ns}
\end{table}

\begin{figure}
\includegraphics[scale=1]{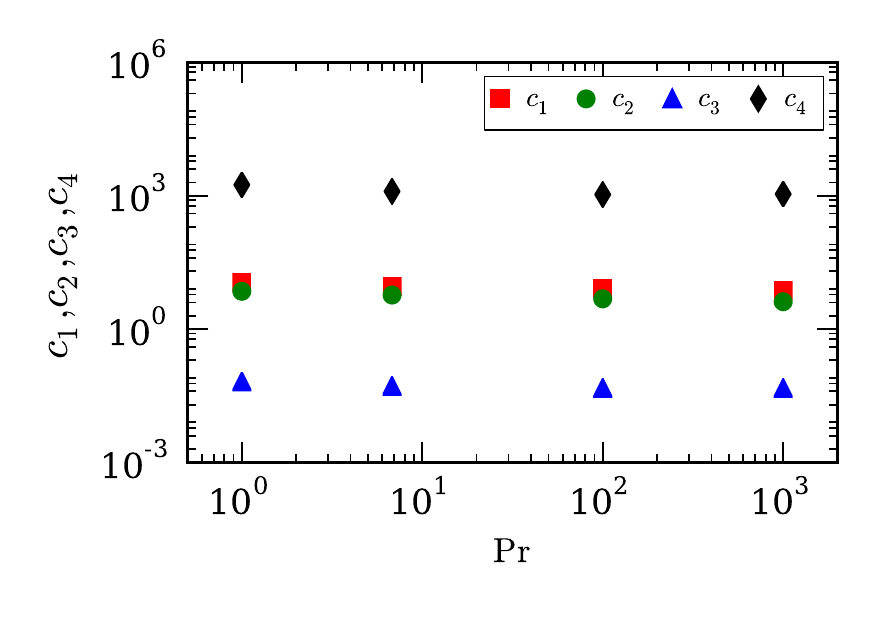}
\caption{For the no-slip boundary condition, the variation of $c_i$'s with $\mathrm{Pr}$ for $\mathrm{Ra} = 2 \times 10^7$. All the coefficients decrease weakly with the increase of the Prandtl number.}
\label{fig:c_pr}
\end{figure}

\begin{figure*}
\includegraphics[scale=0.85]{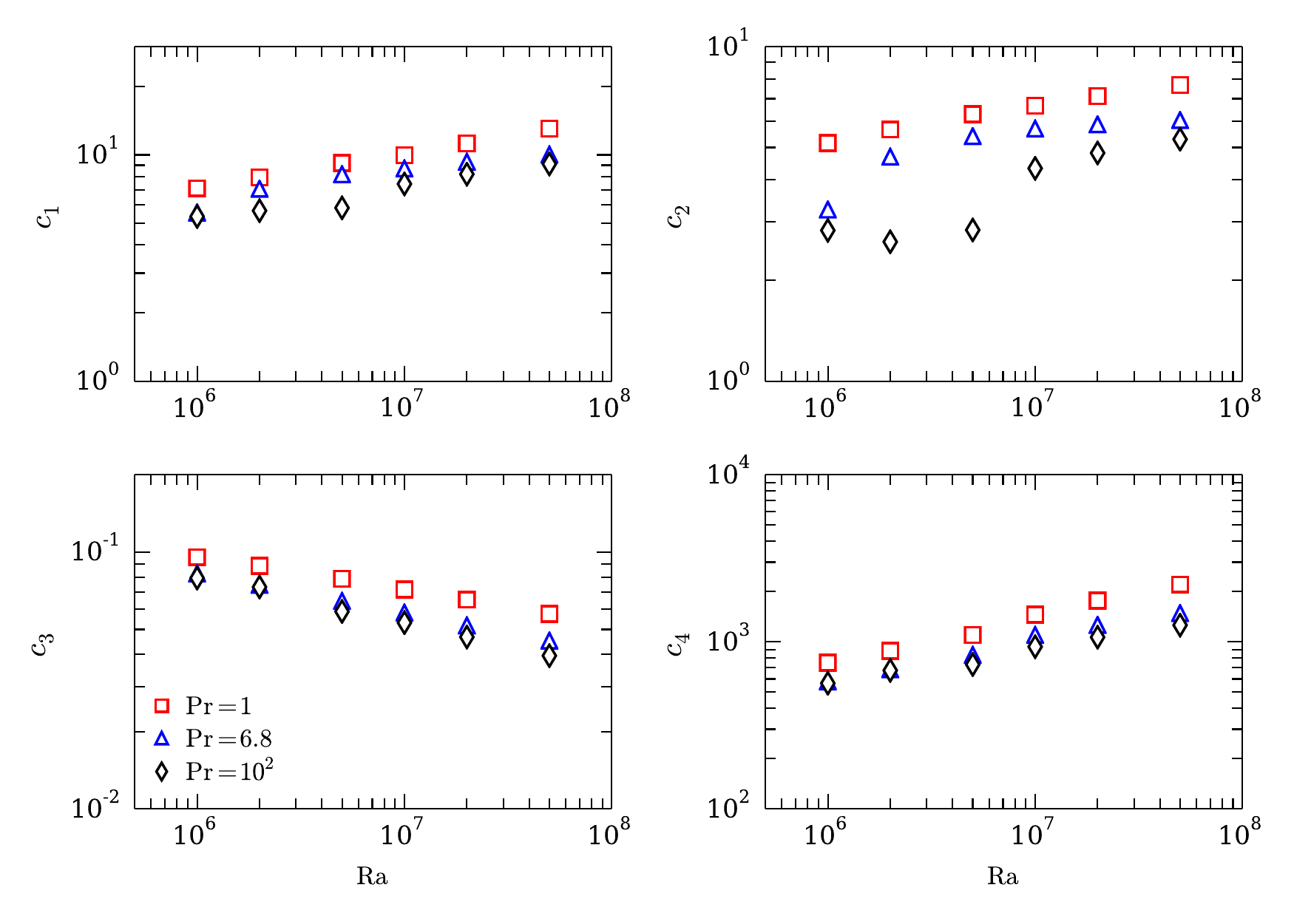}
\caption{For the no-slip boundary condition, the coefficients $c_i$'s as a function of $\mathrm{Ra}$. $c_1, c_2$, and $c_4$ increase with increasing $\mathrm{Ra}$, whereas $c_3$ decreases. The scaling exponents are nearly the same for all the Prandtl numbers. Legend applies to all the plots.}
\label{fig:c_ra}
\end{figure*}

Using the rms values of the above quantities we deduce that the functional dependence of $c_i$'s are of the form listed in Table~\ref{table:c_ns}. Following the similar approach as by Lam \textit{et al.}~\cite{Lam:PRE2002} and Xia \textit{et al.}~\cite{Xia:PRL2002} to determine the functional dependence of $\mathrm{Re(Ra,Pr)}$ and $\mathrm{Nu(Ra,Pr)}$ respectively, we first determine the $\mathrm{Ra}$ dependence of $c_i$'s for $\mathrm{Pr} = 1, 6.8$, and $10^2$ and find that the scaling exponents are nearly similar for these Prandtl numbers. Then we determine the $\mathrm{Pr}$ dependence of $c_i$'s for $\mathrm{Ra} = 2 \times 10^7$. Combining these results, we obtain the functional dependence of $c_i$'s, which are listed in Table~\ref{table:c_ns};  the errors in the exponents of $c_i$'s are $\lessapprox 0.01$, except for the $c_4-\mathrm{Ra}$ scaling where the error is approximately 0.1. We also obtain nearly the same prefactors and exponents by fitting the coefficients with the least square method. Clearly, $c_i$'s are weak functions of $\mathrm{Pr}$, but their dependence on $\mathrm{Ra}$ are reasonably strong so as to affect the  $\mathrm{Pe}$ scaling significantly. Please note that the exponents of $c_i-\mathrm{Ra}$ scaling depend weakly on the Prandtl number. Therefore the exponents in Table~\ref{table:c_ns} are chosen as the average exponent for all the Prandtl numbers.  In Fig.~\ref{fig:c_pr}, we plot $c_i$'s as function of $\mathrm{Pr}$ for $\mathrm{Ra} = 2 \times 10^7$ that exhibits approximately constant values. In Fig.~\ref{fig:c_ra}, we exhibit the variation of $c_i$'s with $\mathrm{Ra}$ for $\mathrm{Pr} = 1, 6.8$, and $10^2$.    

\begin{figure}
\includegraphics[scale=1.2]{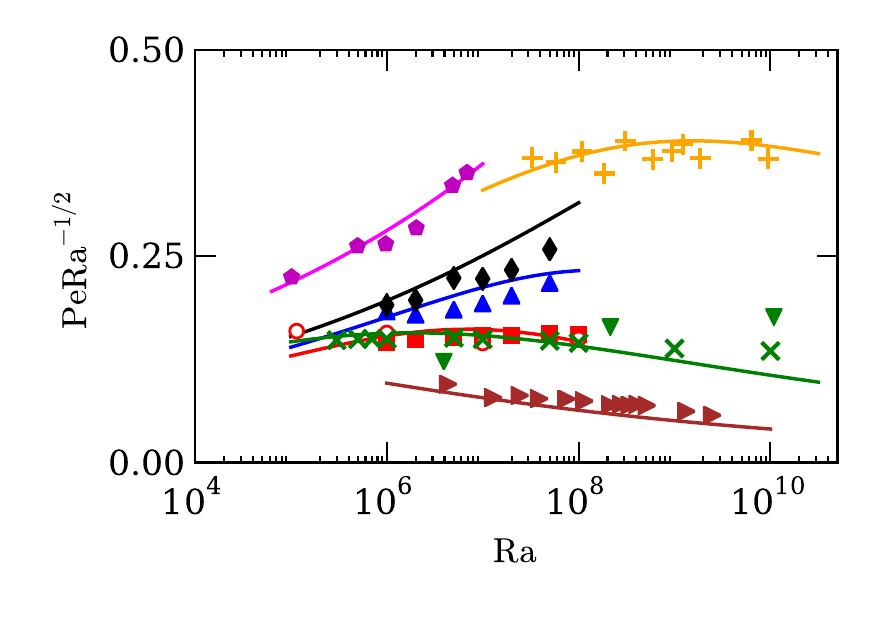}
\caption{For the no-slip boundary condition: the normalized P\'{e}clet number ($\mathrm{PeRa^{-1/2}}$) vs. $\mathrm{Ra}$ for our numerical data for $\mathrm{Pr} = 1$ (red squares), $\mathrm{Pr} = 6.8$ (blue triangles), and $\mathrm{Pr} = 10^2$ (black diamonds); numerical data of Silano \textit{et al.}~\cite{Silano:JFM2010} (magenta pentagons, $\mathrm{Pr} = 10^3$), Reeuwijk \textit{et al.}~\cite{Reeuwijk:PRE2008} (red circles, $\mathrm{Pr} = 1$), Scheel and Schumacher~\cite{Scheel:JFM2014} (green crosses, $\mathrm{Pr} = 0.7$); and the experimental data of Xin and Xia~\cite{Xin:PRE1997} (orange pluses, $\mathrm{Pr} \approx 6.8$), Cioni \textit{et al.}~\cite{Cioni:JFM1997} (brown right triangles, $\mathrm{Pr} \approx 0.022$), and Niemela \textit{et al.}~\cite{Niemela:JFM2001} ($\mathrm{Pr} \approx 0.7$, green down-triangles). The continuous curves represent $\mathrm{Pe}$ computed using our model [Eq.~(\ref{eq:Pe_analy})].}
\label{fig:pe_ns}
\end{figure}

In Fig.~\ref{fig:pe_ns}, we plot the normalized P\'{e}clet number, $\mathrm{PeRa^{-1/2}}$, computed using our simulation data for $\mathrm{Pr} = 1, 6.8, 10^2$. The figure also contains the numerical results of Silano \textit{et al.}~\cite{Silano:JFM2010} ($\mathrm{Pr} = 10^3$), Reeuwijk \textit{et al.}~\cite{Reeuwijk:PRE2008} ($\mathrm{Pr} = 1$), Scheel and Schumacher~\cite{Scheel:JFM2014} ($\mathrm{Pr} = 0.7$), and  the experimental results of Xin and Xia~\cite{Xin:PRE1997} (water, $\mathrm{Pr} \approx 6.8$), Cioni \textit{et al.}~\cite{Cioni:JFM1997} (mercury, $\mathrm{Pr} \approx 0.022$), and Niemela \textit{et al.}~\cite{Niemela:JFM2001} (helium, $\mathrm{Pr} \approx 0.7$). The continuous curves of Fig.~\ref{fig:pe_ns} are the analytically computed $\mathrm{Pe}$ using Eq.~(\ref{eq:Pe_analy}) with the coefficients $c_i$'s listed in Table~\ref{table:c_ns}.  We observe that the theoretical predictions of Eq.~(\ref{eq:Pe_analy}) match quite well with the numerical and experimental results, thus exhibiting usefulness of the model. The predictions of Eq.~(\ref{eq:Pe_analy}) for $\mathrm{Pr} = 0.022$ and $\mathrm{Pr} = 6.8$ have been multiplied with 2.5 and 1.2, respectively, to fit the experimental results from Cioni \textit{et al.}~\cite{Cioni:JFM1997} and Xin and Xia~\cite{Xin:PRE1997}.  The correspondence between our predictions and the past experimental and numerical results shows that $\mathrm{Pe}$ is function of $\mathrm{Pr}$ and $\mathrm{Ra}$, and it depends weakly on geometrical factor and aspect ratio.   Cioni \textit{et al.}~\cite{Cioni:JFM1997} and Xin and Xia~\cite{Xin:PRE1997} performed their experiments on cylinder, while our prediction is for a cube.   Hence, the multiplication factors of 2.5 and 1.2 could be due the aforementioned geometrical factor. 

Using  $c_i$'s and Eq.~(\ref{eq:Ra_cond_turb}) we deduce that $\mathrm{Ra} \ll 10^6 \mathrm{Pr}$  belong to the viscous regime and $\mathrm{Ra} \gg 10^6 \mathrm{Pr}$  belong to the turbulent regime.   For $\mathrm{Pr} = 100$, the $\mathrm{Ra}$ in our simulations belong to this regime, for which our formula predicts
\begin{equation}
\mathrm{Pe} = \frac{c_3}{c_4} \mathrm{Ra} \approx  0.038 \mathrm{Ra}^{0.60}.
\end{equation}	
Our model prediction of $\mathrm{Pe}$ is approximately independent of $\mathrm{Pr}$, and it  is consistent with the results of Silano \textit{et al.}~\cite{Silano:JFM2010}, Horn \textit{et al.}~\cite{Horn:JFM2013},  and Pandey \textit{et al.}~\cite{Pandey:PRE2014}.  Encouraged by this observation, we compare our theoretical predictions with the observations of  Earth's mantle for which $\mathrm{Pr} \gg 1$. The parameters for the mantle are~\cite{Schubert:book2001, Turcotte:book2002, Galsa:SE2015} $d \approx 2900 \mathrm{km}$, $\kappa \approx 10^{-6} \mathrm{m^2/s}$, $\mathrm{Pr} \approx 10^{23}-10^{24}$, $\mathrm{Ra} \approx 5 \times 10^7$, and $U \approx 2$ cm/yr that yields $\mathrm{Pe}_\mathrm{est.} \approx 1840$.  For these parameters,  Eq.~(\ref{eq:Pe_analy}) predicts $\mathrm{Pe}_\mathrm{model} \approx 1580$, which is very close to the estimated value.

For the parameters $c_i$'s, the prediction of Eq.~(\ref{eq:Pe_turb}) yields  
\begin{equation}
\mathrm{Pe} = \sqrt{\frac{c_3}{|c_1-c_2|}} \sqrt{\mathrm{RaPr}} \approx \sqrt{\mathrm{7.5 Pr}} \mathrm{Ra}^{0.38}.
\label{eq:Pe_turb_0.38}
\end{equation}	
Cioni \textit{et al.}~\cite{Cioni:JFM1997} observed that the Reynolds number scales as $\mathrm{Re \sim Ra^{0.424}}$ for $\mathrm{Pr} = 0.025$, which is near our predicted exponent of 0.38.   According to the model estimates, the range of $\mathrm{Ra}$ of Cioni \textit{et al.}~\cite{Cioni:JFM1997}, $5 \times 10^6 \le \mathrm{Ra} \le 5 \times 10^9$,  belongs to  the turbulent regime.  Hence our results are in good agreement with the experimental results of Cioni \textit{et al.}~\cite{Cioni:JFM1997}.  Interestingly, the predicted exponent  of 0.38 for the turbulent regime is quite close to the predictions of $2/5$ by Grossmann and Lohse~\cite{Grossmann:JFM2000} for regime II, which is dominated by $\epsilon_{u,bulk}$ and $\epsilon_{T, BL}$. Here $\epsilon_{u,bulk}$ refers to the kinetic dissipation rate in the bulk, while $\epsilon_{T, BL}$ refers to the thermal dissipation rate in the boundary layer.

Our numerical results for $\mathrm{Pr}=1$ and those of Verzicco and Camussi~\cite{Verzicco:JFM1999},  Reeuwijk \textit{et al.}~\cite{Reeuwijk:PRE2008}, and Niemela \textit{et al.}~\cite{Niemela:JFM2001} yield $\mathrm{Pe} \sim \mathrm{Ra}^{1/2}$, which differs from the predictions of Eq.~(\ref{eq:Pe_turb_0.38}).   It may be due to the fact that  our data for $\mathrm{Pr} = 1$ do not clearly satisfy the inequality $\mathrm{Ra} \gg 10^6 \mathrm{Pr}$.  The data for $\mathrm{Pr = 6.8}$ lie at the boundary between the two regimes, and those for $\mathrm{Pr} = 10^2$ are in viscous regime. The Rayleigh numbers in the experiment of Niemela \textit{et al.}~\cite{Niemela:JFM2001} are very high, hence we expect Eq.~(\ref{eq:Pe_turb_0.38}) to hold instead of $\mathrm{Pe} \sim \mathrm{Ra}^{1/2}$.  The discrepancy between the model prediction and Niemela \textit{et al.}'s~\cite{Niemela:JFM2001} experimental exponent may be due to the fact the experimental $U$ was measured by probes near the wall of the cylinder, which is not same as the volume average assumed in the derivation of Eq.~(\ref{eq:Pe_turb_0.38}).

In the next section, we will discuss the scaling of the Nusselt number and the dissipation rates.

\section{Scaling of viscous term, Nusselt number, and dissipation rates}
\label{sec:nu_diss}

The dependence of $c_i$'s on $\mathrm{Ra}$ and $\mathrm{Pr}$, which is due to the wall effects, affects the scaling of other bulk quantities, e.g., dissipation rates, Nusselt number, etc.  We list some of the effects below.

\subsection{Reynolds number revisited}
\label{subsec:Re}
For an unbounded or free turbulence, the ratio of the nonlinear term, ${\bf u} \cdot \nabla {\bf u}$, and the viscous term is the Reynolds number $Ud/\nu$. But this is not the case for RBC. The ratio
\begin{equation}
\frac{\mathrm{Nonlinear~term}}{\mathrm{Viscous~term}} = \frac{|{\bf u \cdot \nabla u}|}{|\nu \nabla^2 {\bf u}|} =  \frac{Ud}{\nu} \frac{c_1}{c_4} \sim \mathrm{Re} \mathrm{Ra}^{-0.14}.
\end{equation}
Thus, for the same $U, L$, and $\nu$, RBC has a weaker nonlinearity compared to the free or unbounded turbulence.  This effect is purely due to the walls or the boundary layers.   

\subsection{Nusselt number scaling}

\begin{figure}
\includegraphics[scale=1.2]{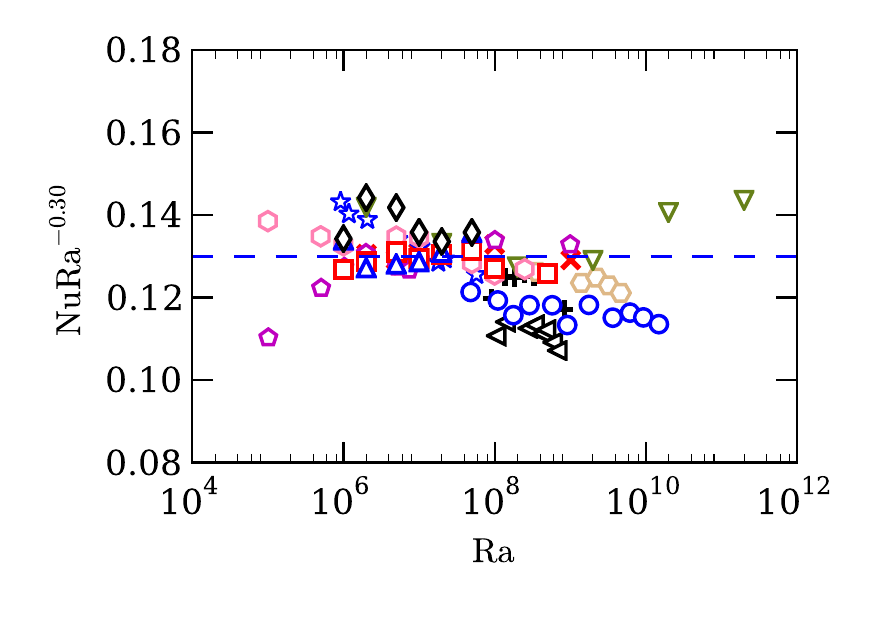}
\caption{For the no-slip boundary condition: the normalized Nusselt number ($\mathrm{NuRa^{-0.30}}$) vs.~$\mathrm{Ra}$. Experimental data: Xia {\it et al.}~\cite{Xia:PRL2002} ($\mathrm{Pr}$ = 205, black pluses; $\mathrm{Pr}$ = 818, black left-pointing triangles), Cioni {\it et al.}~\cite{Cioni:JFM1997} ($4 < \mathrm{Pr} < 8.6$, burlywood horizontal hexagons), Zhou {\it et al.}~\cite{Zhou:JFM2012} ($5.2 < \mathrm{Pr} < 7$, blue stars), Xin and Xia~\cite{Xin:PRE1997} ($\mathrm{Pr} \simeq 7$, blue circles). Numerical data: our data ($\mathrm{Pr} = 1$, red squares; $\mathrm{Pr} = 6.8$, blue triangles; $\mathrm{Pr} = 10^2$, black diamonds), Silano {\it et al.}~\cite{Silano:JFM2010} ($\mathrm{Pr} = 10^2$, red crosses and $\mathrm{Pr} = 10^3$, magenta pentagons), Stevens \textit{et al.}~\cite{Stevens:JFM2010} ($\mathrm{Pr}$ = 0.7, downward green triangles), and Scheel \textit{et al.}~\cite{Scheel:JFM2012} ($\mathrm{Pr}$ = 0.7, pink vertical hexagons). The dashed line represents $\mathrm{Nu} = 0.13\mathrm{Ra}^{0.30}$.}
\label{fig:nu_ns}
\end{figure}

In RBC, the flow is anisotropic due to the presence of buoyancy, which leads to a convective heat transport, quantified using Nusselt number~\cite{Ahlers:RMP2009, Chilla:EPJE2012, Xia:TAML2013}, as
\begin{equation}
\mathrm{Nu} = \frac{\kappa \Delta/d + \langle u_z \theta_{\mathrm{res}} \rangle_V}{\kappa \Delta/d} = 1 + \left\langle \frac{u_z d}{\kappa} \frac{\theta_{\mathrm{res}}}{\Delta} \right\rangle_V = 1 + C_{u\theta_\mathrm{res}} \langle u_z^{'2} \rangle_V^{1/2} \langle \theta_\mathrm{res}^{'2} \rangle_V^{1/2}, \label{eq:Nu}
\end{equation}
where $\langle \rangle_V$ stands for a volume average,  $u'_z =  u_z d /\kappa$, $\theta'_{\mathrm{res}} = \theta_{\mathrm{res}}/\Delta$, and  the normalized correlation function between the vertical velocity and the residual temperature fluctuation~\cite{Verma:PRE2012} is
\begin{equation}
C_{u\theta_\mathrm{res}} = \frac{\langle u'_z \theta'_\mathrm{res} \rangle_V} {\langle u_z^{'2} \rangle_V^{1/2} \langle \theta_\mathrm{res}^{'2} \rangle_V^{1/2}}.
\end{equation}
We compute the above quantities using the numerical data for various $\mathrm{Ra}$ and $\mathrm{Pr}$.  In Fig.~\ref{fig:nu_ns}, we plot the normalized Nusselt number, $\mathrm{NuRa^{-0.30}}$, vs.~$\mathrm{Ra}$ for our results, as well as earlier numerical~\cite{Silano:JFM2010, Stevens:JFM2010, Scheel:JFM2012} and experimental results~\cite{Cioni:JFM1997, Xin:PRE1997, Xia:PRL2002, Zhou:JFM2012}.  The plot indicates that the Nusselt number exponent is close to 0.30, and it is in good agreement with the earlier results for whom the exponents range from 0.27 to 0.33.

The deviation of the exponent from 1/2 (ultimate regime~\cite{Kraichnan:POF1962}) is due to nontrivial correlation $C_{u\theta_\mathrm{res}}$ between $u_z$ and $\langle \theta_\mathrm{res} \rangle$.  In Table~\ref{table:scaling_summary_ns}, we list the scaling of $\mathrm{Nu}$ and $C_{u\theta_{\mathrm{res}}}$ in the turbulent and viscous regimes. The results show that $C_{u\theta_\mathrm{res}} $ and $\langle \theta_\mathrm{res} \rangle$ scale with $\mathrm{Ra}$ in such a way that $\mathrm{Nu} \sim \mathrm{Ra}^{0.32}$, that is primarily due to boundary layer.  Without these corrections, in the turbulent regime, $\mathrm{Nu} \sim \mathrm{Ra}^{1/2}$, as predicted by Kraichnan~\cite{Kraichnan:POF1962}. Lohse and Toschi~\cite{Lohse:PRL2003} performed numerical simulation of RBC with periodic boundary condition, and showed that $\theta_\mathrm{res} \sim \Delta$ and $\langle u_z \theta_\mathrm{res}  \rangle \sim \mathrm{Ra}^{1/2}$ in the absence of any boundary.  He {\em et al.}~\cite{He:PRL2012} argued that the boundary layer becomes turbulent at $\mathrm{Ra} \sim 10^{15}$.  Hence $\langle u_z \theta_\mathrm{res}  \rangle$ may start to show $\mathrm{Ra}^{1/2}$ scaling, as indicated by He {\em et al.}~\cite{He:PRL2012}, which will occur when $C_{u\theta_\mathrm{res}}$ will become independent of $\mathrm{Ra}$.    

\begin{table}
\caption{Scaling of the correlation function $C_{u\theta_\mathrm{res}}$, $\langle \theta_\mathrm{res}^2 \rangle^{1/2}$, $\langle u_z^2 \rangle^{1/2} $, $\mathrm{Nu}$, and the global dissipation rates computed using numerical data for the no-slip boundary condition. The errors in the exponents of $C_{u\theta_\mathrm{res}}$ and $\langle \theta_\mathrm{res}^2 \rangle^{1/2}$ are approximately $0.01$, and those of the other quantities are approximately $0.02$.}
\begin{ruledtabular}
\begin{tabular}{ccc}
	& Turbulent regime & Viscous regime \\
\hline 
$C_{u \theta_\mathrm{res}}$ &  $\mathrm{Ra}^{-0.05}$   & $\mathrm{Ra}^{-0.07}$ \\
$\langle \theta_\mathrm{res}^2 \rangle^{1/2} $  & $\mathrm{Ra}^{-0.13}$ &  $\mathrm{Ra}^{-0.18}$\\
$\langle u_z^2 \rangle^{1/2} $  & $\mathrm{Ra}^{0.51}$ &  $\mathrm{Ra}^{0.58}$ \\ 
$\mathrm{Nu}$  & $\mathrm{Ra}^{0.32}$ &  $\mathrm{Ra}^{0.33}$ \\ 
$\epsilon_u$ & $(U^3/d) \mathrm{Ra}^{-0.21}$ &  $(\nu U^2/d^2)   \mathrm{Ra}^{0.15}$ \\
$\epsilon_T$ & $(U \Delta^2/d) \mathrm{Ra}^{-0.19}$ &  $(U \Delta^2/d)  \mathrm{Ra}^{-0.25}$ \\
\end{tabular}
\end{ruledtabular}
\label{table:scaling_summary_ns}
\end{table}

\subsection{Scaling of dissipation rates}
The kinetic energy supplied by the buoyancy is dissipated by the viscous forces.  Shraiman and Siggia~\cite{Shraiman:PRA1990} derived that the viscous dissipation rate, $\epsilon_u$, is
\begin{equation}
\epsilon_u = \nu |\nabla \times {\bf u}|^2 =   \frac{\nu^3}{d^4} \frac{\mathrm{(Nu-1)Ra}}{\mathrm{Pr}^2} =  \frac{U^3}{d}  \frac{\mathrm{(Nu-1)RaPr}}{\mathrm{Pe}^3}.
\end{equation}
In the turbulent regime of our simulation, $ \mathrm{Nu} \sim \mathrm{Ra}^{0.32}$ and $\mathrm{Pe} \sim \sqrt{\mathrm{Ra}}$, hence, $\epsilon_u  \ne U^3/d$, rather 
\begin{equation}
\epsilon_u \sim \frac{U^3}{d}  \mathrm{Ra}^{-0.21}.
\label{eq:epsilon_u_m0.2}
\end{equation}
The viscous dissipation rate, which is equal to the energy flux, is smaller than $U^3/d$ due to weaker nonlinearity compared to the unbounded flows (see Sec.~\ref{subsec:Re}); this is due to the boundary layers. 

In the viscous regime,
\begin{equation}
\epsilon_u  = \frac{\nu U^2}{d^2}  \frac{\mathrm{(Nu-1)Ra}}{\mathrm{Pe}^2}.
\end{equation}
Since  $ \mathrm{Nu} \sim \mathrm{Ra}^{0.33}$ and $ \mathrm{Pe} \sim \mathrm{Ra}^{0.58}$, we observe that 
\begin{equation}
\epsilon_u= \frac{\nu U^2}{d^2}  \mathrm{Ra}^{0.17}.
\label{eq:epsilon_u_0.1}
\end{equation}
Thus, RBC has a larger $\epsilon_u$ compared to unbounded flows due to boundary layers.   

Similar results follow for the thermal dissipation rate, $\epsilon_T$.  According to one of the exact relations of Shraiman and Siggia~\cite{Shraiman:PRA1990}
\begin{equation}
\epsilon_T =  \kappa |\nabla T|^2 =  \kappa \frac{\Delta^2}{d^2} \mathrm{Nu} =  \frac{U \Delta^2}{d} \frac{\mathrm{Nu}}{\mathrm{Pe}}.
\end{equation}
For both the turbulent and viscous regimes we employ $\epsilon_T \approx U \theta^2/d \approx U \Delta^2/d$ since the nonlinear term dominates the diffusion term in the temperature equation.  This is because $\mathrm{Pe} \gg 1$ for all our runs.

Hence, substitution of the expressions for $\mathrm{Pe}$ and $\mathrm{Nu}$ in the above equation yields the following $\epsilon_T $ for the turbulent regime of our simulations:
 \begin{equation}
\epsilon_T \sim \frac{U \Delta^2}{d}  \mathrm{Ra}^{-0.19},
\label{eq:eps_T1}
\end{equation}
but 
\begin{equation}
\epsilon_T \sim    \frac{U \Delta^2}{d}  \mathrm{Ra}^{-0.25}
\label{eq:eps_T2}
\end{equation}
for the viscous regime.  The above $\mathrm{Ra}$-dependent corrections are also due to the boundary layers.  In the turbulent regime, for $\mathrm{Pr} = 1$, the ratio of the nonlinear term of the temperature equation and the thermal diffusion term is
\begin{equation}
\frac{{\bf u \cdot} \nabla \theta }{\kappa \nabla^2 \theta } = \frac{c_5}{c_6} \frac{Ud}{\kappa} \sim  \mathrm{Ra}^{-0.30} \mathrm{Pe},
\end{equation}
since 
\begin{eqnarray}
c_5 & = & \frac{|{\bf u \cdot \nabla \theta}|} {U \theta /d} \sim \mathrm{Ra}^{0.09}, \nonumber \\
c_6 & = & \frac{|\nabla^2 \theta|} {\theta/d^2} \sim \mathrm{Ra}^{0.39}.
\end{eqnarray}
Thus, the nonlinearity in the temperature equation [Eq.~(\ref{eq:th})] of RBC is weaker than the corresponding term in unbounded flow (e.g., passive scalar in a periodic box).  Consequently the entropy flux is weaker than that for unbounded flows, which is the reason for the behavior of Eqs.~(\ref{eq:eps_T1},\ref{eq:eps_T2}).

\begin{figure}
\includegraphics[scale=1]{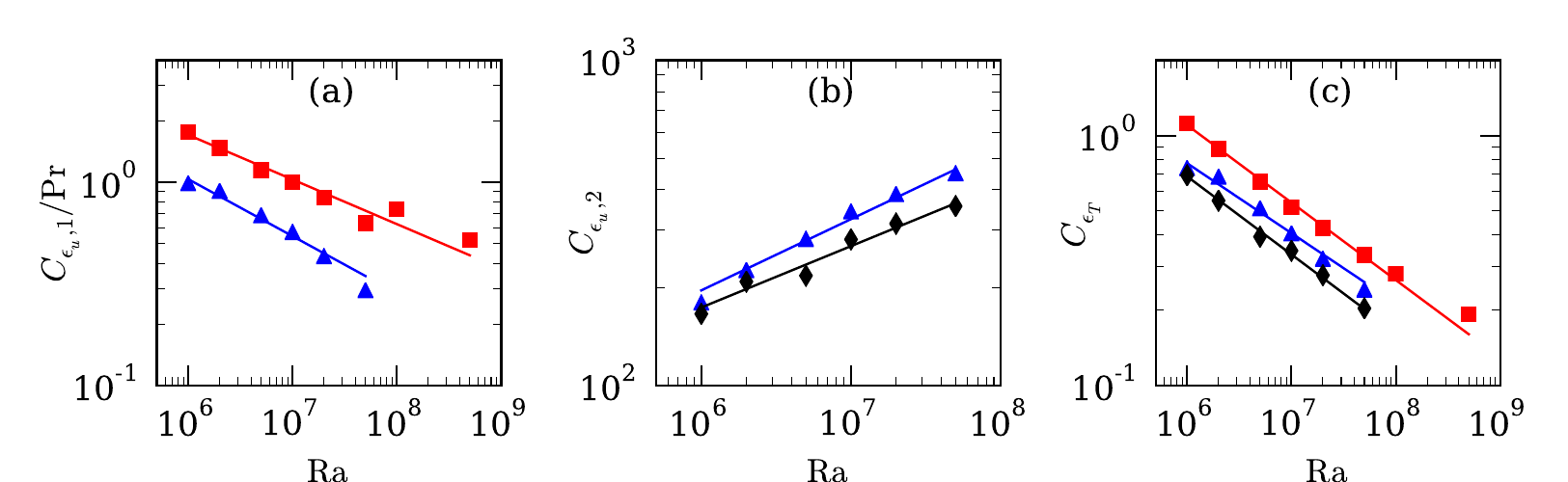}
\caption{The normalized dissipation rates for the no-slip boundary condition: (a) $C_{\epsilon_u,1}/\mathrm{Pr}$, (b) $C_{\epsilon_u,2}$, and (c) $C_{\epsilon_T}$ as functions of $\mathrm{Ra}$ for $\mathrm{Pr} = 1$ (red squares), $\mathrm{Pr} = 6.8$ (blue triangles), and $\mathrm{Pr} = 10^2$ (black diamonds). The best fits to the data are depicted as solid lines.}
\label{fig:c_eps}
\end{figure}

We numerically compute the following normalized dissipation rates:   
\begin{eqnarray}
C_{\epsilon_u,1} & = &  \frac{\epsilon_u}{U^3/d} = \frac{\mathrm{(Nu-1)RaPr}}{\mathrm{Pe}^3} \sim \mathrm{Ra}^{-0.21} \mathrm{Pr},~~\mathrm{(turbulent~regime)}  \label{eq:Cueps1}\\
C_{\epsilon_u,2} & = & \frac{\epsilon_u}{\nu U^2/d^2} = \frac{\mathrm{(Nu-1)Ra}}{\mathrm{Pe}^2} \sim \mathrm{Ra} ^{0.17}, ~~\mathrm{(viscous~regime)}  \label{eq:Cueps2} \\
C_{\epsilon_T} & = &  \frac{\epsilon_T}{U \theta^2/d}=   \frac{\mathrm{Nu}}{\mathrm{Pe}} \sim \mathrm{Ra} ^{-0.25}, \label{eq:ct}
\end{eqnarray}
which are plotted in Fig.~\ref{fig:c_eps}.  We observe that $C_{\epsilon_u,1}/\mathrm{Pr} \sim \mathrm{Ra}^{-0.22 \pm 0.02}$ and $\mathrm{Ra}^{-0.25 \pm 0.03}$ for $\mathrm{Pr} = 1$ and 6.8 respectively, which is in good agreement with Eq.~(\ref{eq:Cueps1}). The exponents for  $C_{\epsilon_u,2}$ are $0.22 \pm 0.01$ and $0.19 \pm 0.02$ for $\mathrm{Pr} = 6.8$ and $10^2$ respectively with reasonable accordance with Eq.~(\ref{eq:epsilon_u_0.1}) for $\mathrm{Pr}=10^2$.  For the thermal dissipation rate, we observe $C_{\epsilon_T} \sim \mathrm{Ra}^{-0.32 \pm 0.02}$ scaling for $\mathrm{Pr} = 1, 6.8$, and $10^2$ consistent with the above scaling.  Table~\ref{table:scaling_summary_ns} lists the $\mathrm{Ra}$-dependence of the dissipation rates in the turbulent and viscous regimes. 

We estimate the dissipation rate (product of the dissipation rate and the appropriate volume) in the bulk, $D_{u,bulk}$, and in the boundary layer, $D_{u,BL}$.  Their ratio is
\begin{eqnarray}
\frac{D_{u,BL}}{D_{u,bulk}} & \approx & \frac{(\epsilon_{u,BL}) (2 A \delta_u)}{(\epsilon_{u,bulk}) (A d - 2 A \delta_u)}  \approx  \left( \frac{2 \nu U^2/\delta_u^2}{U^3/d} \right) \frac{\delta_u}{d} \nonumber \\
&\approx &  2\frac{d/\delta_u}{\mathrm{Re}} \approx2 \mathrm{Re}^{-1/2}
\label{eq:Du}
\end{eqnarray}
since $\delta_u/d \sim \mathrm{Re}^{-1/2}$.~\cite{Landau:FM1987} Here $A$ is the area of the horizontal plates, and $\delta_u$ is the thickness of the viscous boundary layers at the top and bottom plates. Since the dissipation takes place near both the plates, we include a factor 2 here.   Note that we do not substitute  the weak $\mathrm{Ra}$ dependence of  Eqs.~(\ref{eq:epsilon_u_m0.2}, \ref{eq:epsilon_u_0.1}) as an approximation. From Eq.~(\ref{eq:Du}) we deduce that $D_{u,BL} \ll D_{u,bulk}$ for large $\mathrm{Re}$.  However in the viscous regime, the boundary layer extends to the whole region ($2 \delta_u \approx d$), hence $D_{u,BL}$ dominates $ D_{u,bulk}$.  

Earlier, Grossmann and Lohse~\cite{Grossmann:JFM2000, Grossmann:PRL2001, Grossmann:PRE2002} worked out the scaling of the Reynolds and Nusselt numbers by invoking the exact relations of Shraiman and Siggia~\cite{Shraiman:PRA1990} and using the fact that the total dissipation is a sum of those in the bulk and in the boundary layers ($D_{u,bulk}$ and $D_{u,BL}$ respectively).  They employed $\epsilon_{u,bulk}  = U^3/d$, $\epsilon_{T,bulk}  = U \Delta^2 /d$, $\epsilon_{u,BL}  =\nu U^2/\delta_u^2$, and $\epsilon_{T,BL}  = \kappa\Delta^2 /\delta_T^2$; and then  equated one of the expressions in the appropriate regimes.  They also employed corrections for large $\mathrm{Pr}$ and small $\mathrm{Pr}$ cases.   Our model discussed in this paper is an alternative to that of GL with an attempt to highlight the anisotropic effects arising due to the boundary layers that yield $\epsilon_u  \ne U^3/d$ and $\epsilon_T  \ne U \Delta^2 /d$.  Note that we report a single formula for $\mathrm{Pe}$ in comparison to the eight expressions of Grossman and Lohse~\cite{Grossmann:JFM2000} for various limiting cases. 

From the above derivation it is apparent  that the boundary layers of RBC have significant effects on the large-scale quantities; consequently  the flow behavior in RBC is very different from the unbounded fluid turbulence for which we employ homogeneous and isotropic formalism.  In particular, for a free turbulence under the isotropy assumption, $\langle u'_z \theta'_\mathrm{res} \rangle_V = 0$, hence the nonzero $C_{u\theta_\mathrm{res}} $ for RBC is purely due to the walls or boundary layers.  To relate to the scaling in the ultimate regime, we  conjecture that $C_{u\theta_\mathrm{res}} $, $\theta'_\mathrm{res} $, and $c_i$'s would become independent of $\mathrm{Ra}$ due to the detachment of the boundary layer, hence  $\mathrm{Nu} \sim \langle u_z^{'2} \rangle^{1/2} \sim  \mathrm{Ra}^{1/2}$, as predicted by Kraichnan~\cite{Kraichnan:POF1962}.  Note that for a nonzero $\mathrm{Nu}$,  $C_{u\theta_\mathrm{res}} $ must be finite, contrary to the predictions for isotropic turbulence for which $C_{u\theta_{\mathrm{res}}} = 0 $.  We need further experimental inputs as well as numerical simulations at very large $\mathrm{Ra}$ to test the above conjecture. 

Here we end our discussion on RBC with no-slip boundary condition.  In the next section we will discuss the scaling relations for RBC with the free-slip boundary condition.

\section{Results of RBC with free-slip boundary condition}
\label{sec:free_slip}

In this section we will study the scaling of P\'{e}clet number under free-slip boundary condition.  Towards this objective we perform RBC simulations with free-slip walls for a set of Prandtl and Rayleigh numbers, and  compute the strengths of the nonlinear term, pressure gradient, buoyancy, and the viscous force, and the corresponding coefficients $c_i$'s defined in Sec.~\ref{sec:Pe}.  After this we  compute the P\'{e}clet number as a function of $\mathrm{Pr}$ and $\mathrm{Ra}$. The procedure is identical to that described for the no-slip boundary condition.

\begin{figure}
\includegraphics[scale=1]{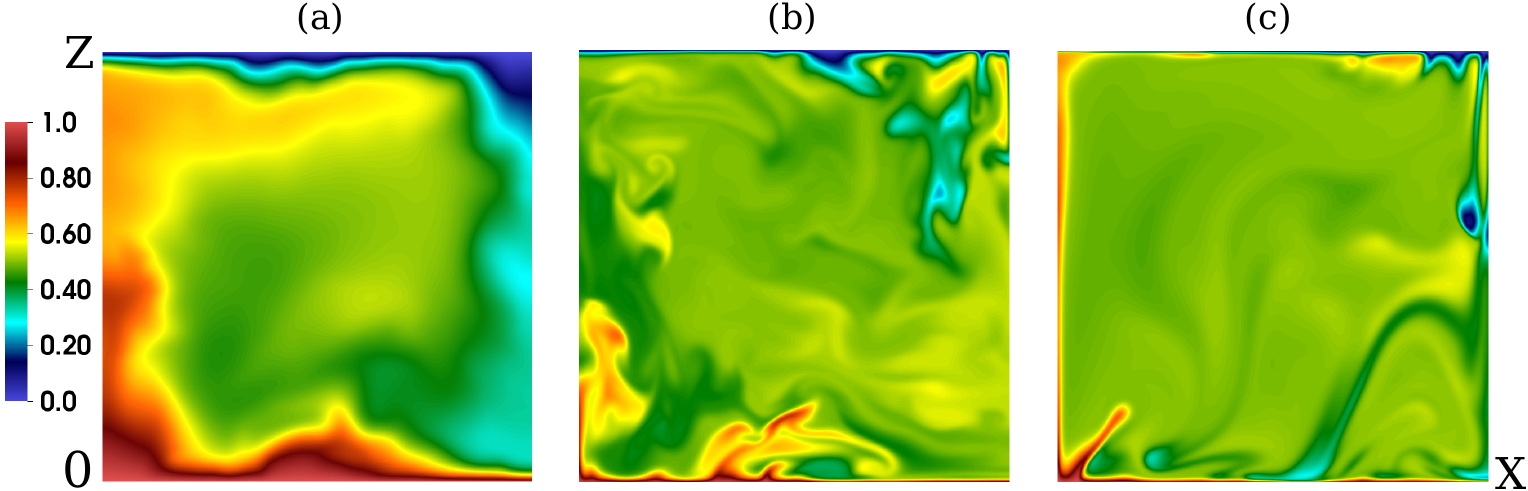}
\caption{For the free-slip boundary condition: the temperature field in a vertical plane at $y = 0.4$ for (a) $\mathrm{Pr} = 0.02, \mathrm{Ra} = 2 \times 10^6$, (b) $\mathrm{Pr} = 1, \mathrm{Ra} = 5 \times 10^7$, and (c) $\mathrm{Pr} = 10^2, \mathrm{Ra} = 5 \times 10^7$. Thermal structures become sharper with increasing Prandtl number.}
\label{fig:temp_field_sf}
\end{figure}

We perform direct numerical simulations for $\mathrm{Pr} = 0.02, 1, 4.38, 10^2, 10^3$, and $\infty$, and Rayleigh numbers between $10^5$ and $2 \times 10^8$ in a three-dimensional unit box using a pseudo-spectral code {\scshape Tarang}~\cite{Verma:Pramana2013}. For the velocity field, we employ free-slip boundary condition at all the walls, and for the temperature field, the  isothermal condition at the top and bottom plates and the adiabatic condition at the vertical walls. We use the fourth-order Runge-Kutta (RK4) method for time discretization, and 2/3 rule to dealiase the fields. We start our simulations for lower Ra using random initial values for the velocity and temperature fields, and then take the steady-state fields as the initial condition to simulate for higher Rayleigh numbers. We  employ $64^3$ to $512^3$ grids and ensure that the Kolmogorov ($\eta_u$) and the Batchelor ($\eta_\theta$) lengths are larger than the mean distance between two adjacent grid points for each simulation. The details of simulation parameters are given in Table~\ref{table:details}. Figure~\ref{fig:temp_field_sf} demonstrates the temperature field in a vertical cross-section of the box at $y = 0.4$. The temperature field is diffusive for $\mathrm{Pr} = 0.02$, whereas the field becomes plume-dominated for larger Prandtl numbers~\cite{Chilla:EPJE2012}.

\begin{table}
\begin{ruledtabular}
\caption{Details of our simulations with the stress-free boundary condition. The quantities are same as in Table~\ref{table:details_ns}, except $\eta$ is the Kolmogorov length scale ($\eta_u$) for $\mathrm{Pr} = 0.02$.}
\label{table:details}
\begin{tabular}{cccccc|cccccc}
$\mathrm{Pr}$ & $\mathrm{Ra}$ & $N^3$ & Nu & Pe & $k_{\mathrm{max}}\eta$ & $\mathrm{Pr}$ & $\mathrm{Ra}$ & $N^3$ & Nu & Pe & $k_{\mathrm{max}}\eta$ \\ \hline
0.02 & $1 \times 10^5$ & $256^3$ & 4.93	& $4.02 \times 10^1$ & 4.6 & $10^2$ & $1 \times 10^6$ & $128^3$ & 20.8 & $7.49 \times 10^2$ & 1.9 \\ 
0.02 & $2 \times 10^5$ & $256^3$ & 5.74	& $5.28 \times 10^1$ & 3.7 & $10^2$ & $2 \times 10^6$ & $256^3$ & 29.0 & $1.23 \times 10^3$ & 3.0 \\ 
0.02 & $5 \times 10^5$ & $512^3$ & 7.21	& $7.71 \times 10^1$ & 5.4 & $10^2$ & $5 \times 10^6$ & $256^3$ & 39.2 & $2.15 \times 10^3$ & 2.2 \\ 
0.02 & $1 \times 10^6$ & $512^3$ & 8.65 & $1.01 \times 10^2$ & 4.3 & $10^2$ & $1 \times 10^7$ & $256^3$ & 45.8 & $3.21 \times 10^3$ & 1.7 \\ 
0.02 & $2 \times 10^6$ & $512^3$ & 10.7 & $1.41 \times 10^2$   & 3.4 & $10^2$ & $2 \times 10^7$ & $256^3$ & 58.0 & $5.00 \times 10^3$ & 1.4 \\ 
1 & $1 \times 10^6$ & $64^3$ & 18.5	& $4.68 \times 10^2$  & 3.0 & $10^2$ & $5 \times 10^7$ & $512^3$ & 77.4 & $8.70 \times 10^3$ & 2.0 \\ 
1 & $2 \times 10^6$ & $64^3$ & 21.9	& $6.07 \times 10^2$  & 2.5 & $10^3$ & $1 \times 10^6$ & $256^3$ & 21.5 & $7.54 \times 10^2$ & 2.1 \\
1 & $5 \times 10^6$ & $128^3$ & 28.4 & $8.84 \times 10^2$ & 3.7 & $10^3$ & $2 \times 10^6$ & $256^3$ & 27.1 & $1.16 \times 10^3$ & 1.7 \\
1 & $1 \times 10^7$ & $128^3$ & 32.6 & $1.16 \times 10^3$ & 3.0 & $10^3$ & $5 \times 10^6$ & $256^3$ & 36.0 & $2.04 \times 10^3$ & 1.2 \\
1 & $2 \times 10^7$ & $128^3$ & 39.5 & $1.57 \times 10^3$ & 2.4 & $10^3$ & $1 \times 10^7$ & $512^3$ & 45.3 & $3.09 \times 10^3$ & 2.0 \\
1 & $5 \times 10^7$ & $256^3$ & 49.1 & $2.36 \times 10^3$ & 3.6 & $10^3$ & $2 \times 10^7$ & $512^3$ & 54.1 & $4.29 \times 10^3$ & 1.6 \\
1 & $1 \times 10^8$ & $256^3$ & 60.1 & $3.11 \times 10^3$ & 2.9 & $10^3$ & $5 \times 10^7$ & $512^3$ & 75.2 & $8.03 \times 10^3$ & 1.2 \\
4.38 & $1 \times 10^6$ & $128^3$ & 21.5 & $6.98 \times 10^2$ & 4.1 & $10^3$ & $1 \times 10^8$ & $512^3$ & 91.4 & $1.25 \times 10^4$ & 0.9 \\
4.38 & $2 \times 10^6$ & $128^3$ & 26.4 & $9.65 \times 10^2$ & 1.6 & $\infty$ & $5 \times 10^6$ & $256^3$ & 35.3 & $2.03 \times 10^3$ & 7.0 \\
4.38 & $5 \times 10^6$ & $128^3$ & 33.9 & $1.47 \times 10^3$ & 2.4 & $\infty$ & $1 \times 10^7$ & $256^3$ & 43.6 & $3.10 \times 10^3$ & 5.6 \\
4.38 & $1 \times 10^7$ & $128^3$ & 41.0 & $1.96 \times 10^3$ & 1.9 & $\infty$ & $2 \times 10^7$ & $256^3$ & 54.4 & $4.46 \times 10^3$ & 4.5 \\
4.38 & $2 \times 10^7$ & $256^3$ & 48.5 & $2.61 \times 10^3$ & 3.2 & $\infty$ & $5 \times 10^7$ & $256^3$ & 72.4 & $7.95 \times 10^3$ & 3.3 \\
4.38 & $5 \times 10^7$ & $256^3$ & 62.3 & $3.88 \times 10^3$ & 2.4 & $\infty$ & $1 \times 10^8$ & $256^3$ & 92.3 & $1.33 \times 10^4$ & 2.6 \\
-- & -- & -- & -- & -- &  -- & $\infty$ & $2 \times 10^8$ & $512^3$ & 113 & $1.92 \times 10^4$ & 4.2 \\
\end{tabular}
\end{ruledtabular} 
\end{table}

\begin{figure}
\includegraphics[scale=1.1]{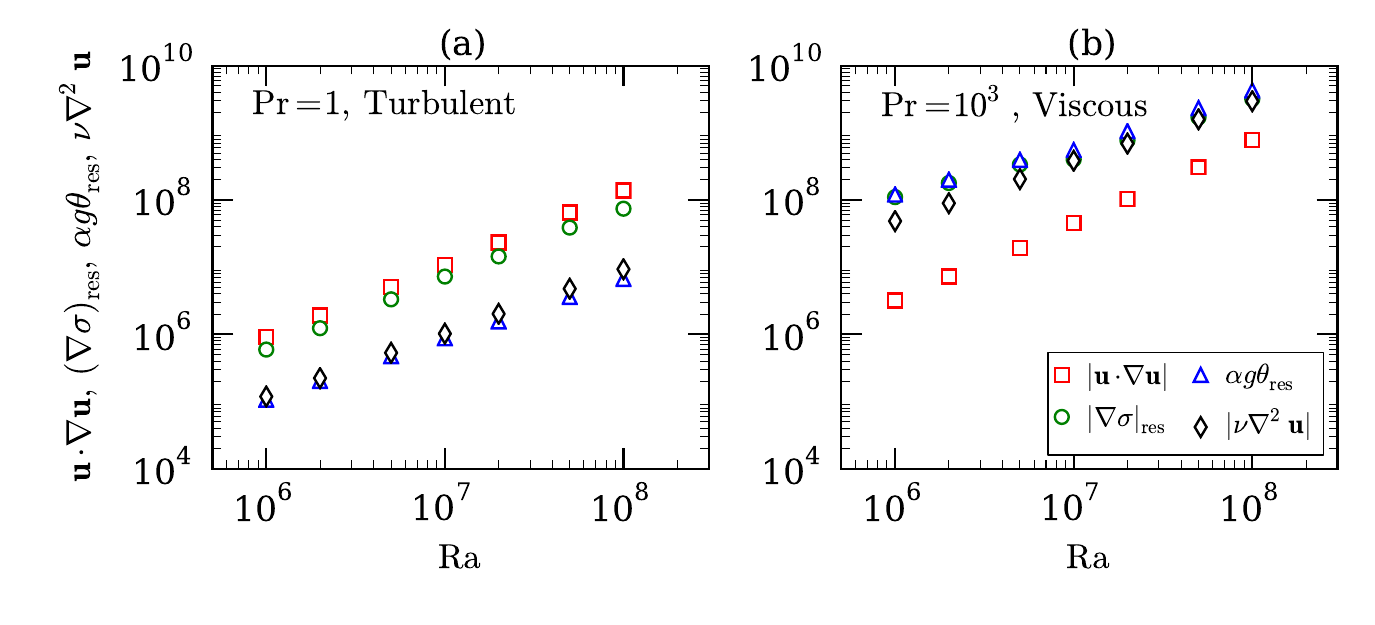}
\caption{For the free-slip boundary condition: comparison of the rms values of ${\bf u} \cdot \nabla {\bf u}$, $(-\nabla \sigma)_{\mathrm{res}}$, $\alpha g \theta_{\mathrm{res}} \hat{\bf z}$, and $\nu \nabla^2 {\bf u}$ as function of $\mathrm{Ra}$ (a) in the turbulent regime ($\mathrm{Pr}=1$), and (b) in viscous regime ($\mathrm{Pr}=10^3$).} 
\label{fig:compare_sf}
\end{figure}

\begin{figure}
\includegraphics[scale=1]{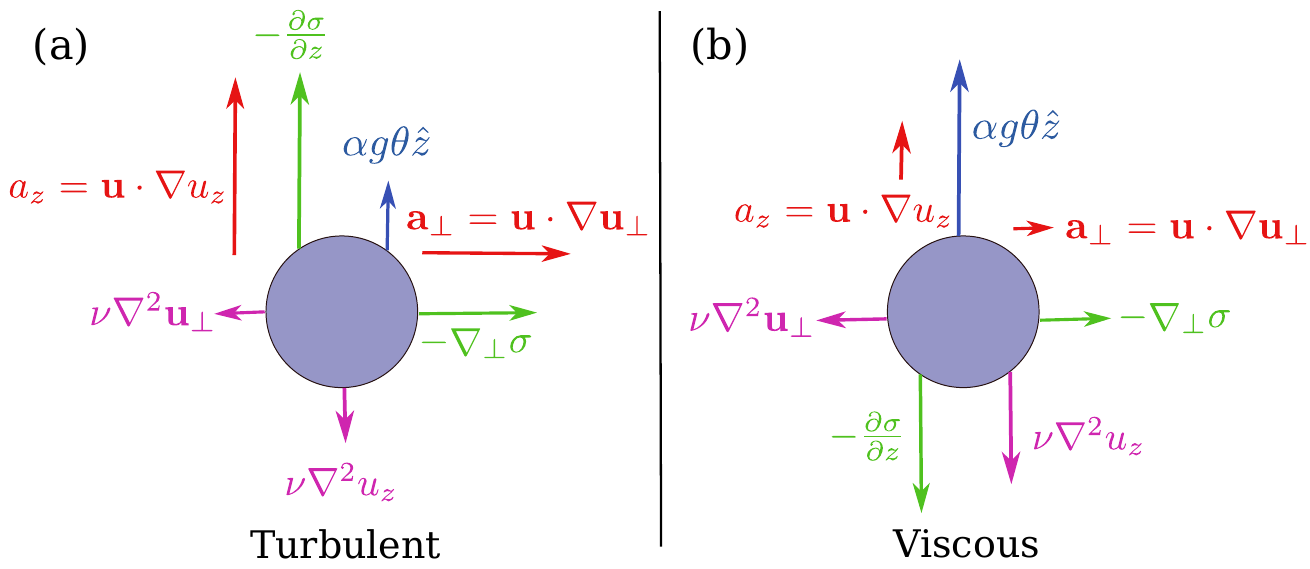}
\caption{For the free-slip boundary condition, the relative strengths of the forces acting on a fluid parcel. In turbulent regime, the acceleration ${\bf u} \cdot \nabla {\bf u}$ is provided primarily by the pressure gradient, both in parallel and perpendicular directions.  In viscous regime, the buoyancy is balanced by the pressure gradient and the viscous force along $\hat{\bf z}$; in the perpendicular direction, the pressure gradient  balances the viscous force.}
\label{fig:schematic_sf}
\end{figure}

We compute the rms values of $|{\bf u \cdot \nabla u}|, |(-\nabla \sigma)_{\mathrm{res}}|, |\alpha g \theta_\mathrm{res} \hat{\bf z}|$, and $|\nu \nabla^2 {\bf u}|$ for $\mathrm{Pr}=1$ and $10^3$.  These values are plotted as function of $\mathrm{Ra}$ in Fig.~\ref{fig:compare_sf}, and their $\mathrm{Ra}$-dependence are given in Table~\ref{table:terms_sf}. From the numerical data we can deduce the following:
\begin{enumerate}
\item In the turbulent regime (for $\mathrm{Pr}=1$ of Fig.~\ref{fig:compare_sf}(a)), the acceleration is dominated by the pressure gradient; the buoyancy and viscous terms are quite weak in comparison.  This feature is same as that for the no-slip boundary condition (see Sec.~\ref{sec:no_slip}).  However, for the free-slip boundary condition, both vertical and horizontal accelerations are significant (see Fig.~\ref{fig:schematic_sf}(a)).

\item In the viscous regime (for $\mathrm{Pr}=10^3$ of Fig.~\ref{fig:compare_sf}(b)), the nonlinear term is weak, and $(-\nabla \sigma)_{\mathrm{res}}, \alpha g \theta_\mathrm{res} \hat{\bf z}$, and $\nu \nabla^2 {\bf u}$ balance each other as shown in  Fig.~\ref{fig:schematic_sf}(b). Interestingly the pressure gradient opposes the motion.  We will revisit this issue in the following discussion.  Note that for the no-slip boundary condition,  the nonlinear term and the pressure gradient are weak (see Sec.~\ref{sec:no_slip}).  
\end{enumerate}

\begin{figure}
\includegraphics[scale=0.9]{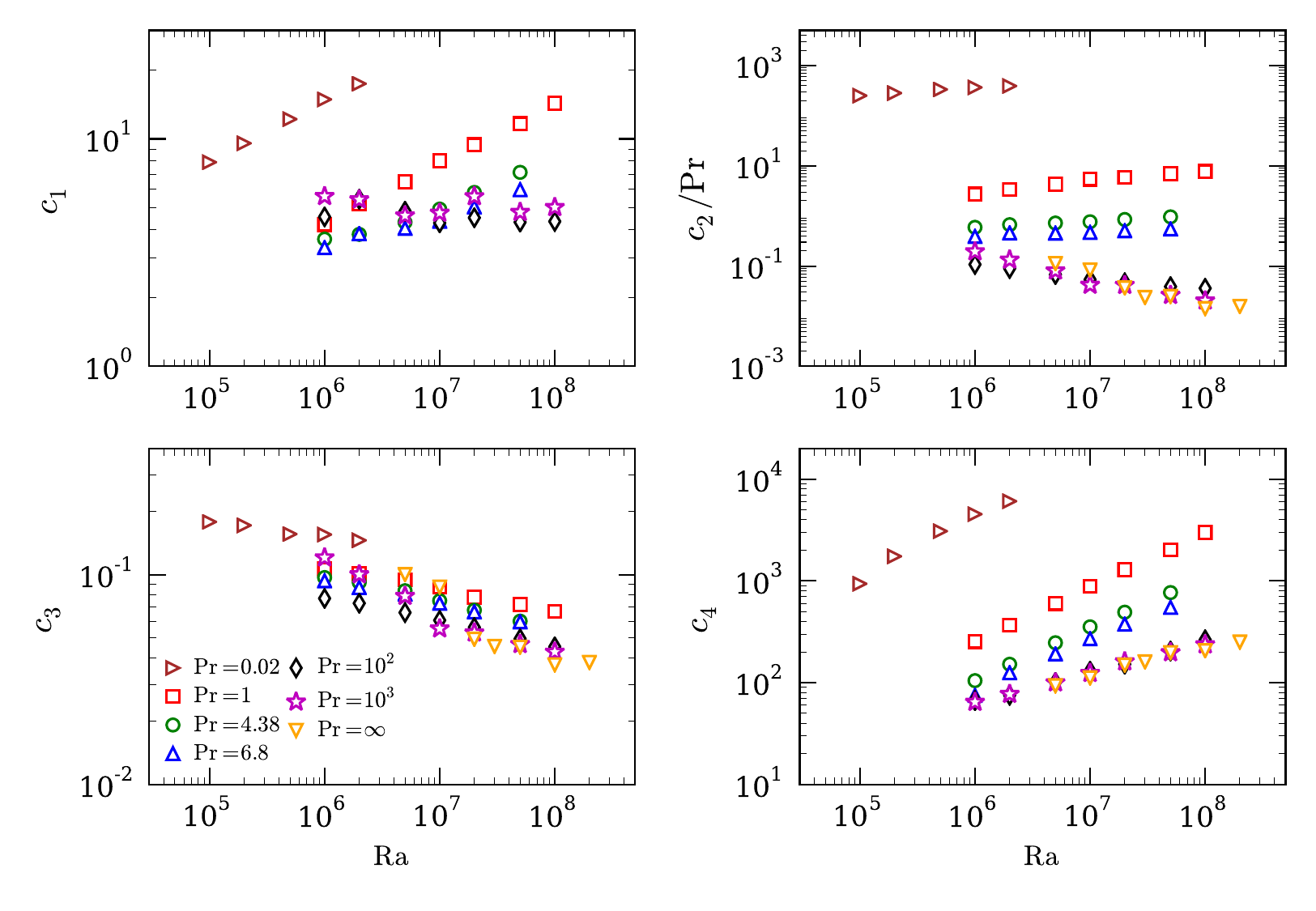}
\caption{For the free-slip boundary condition, the coefficients $c_i$'s as function of $\mathrm{Ra}$. Note that the nonlinear term and consequently the coefficient $c_1$ is zero for $\mathrm{Pr} = \infty$.}
\label{fig:c_ra_sf}
\end{figure}

\begin{figure}
\includegraphics[scale=1]{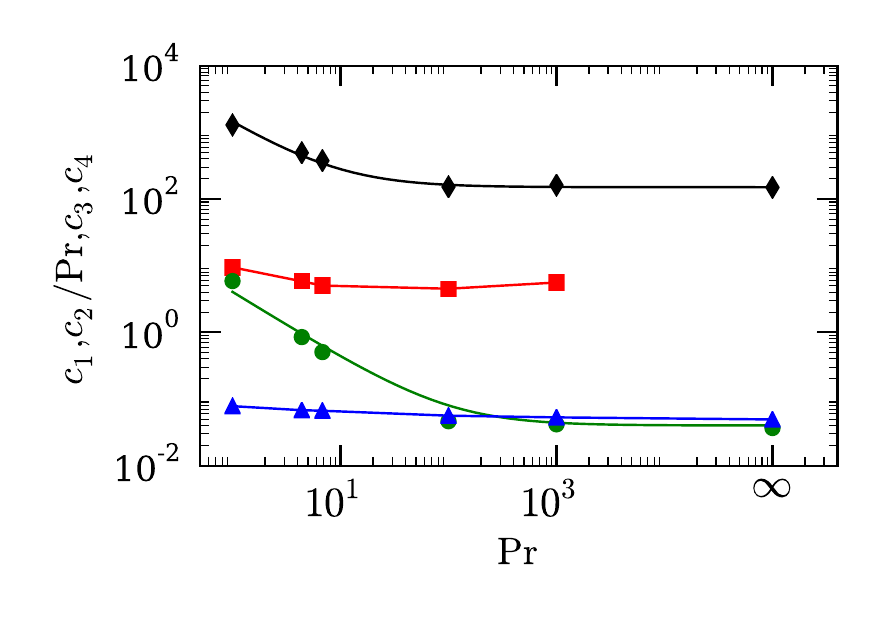}
\caption{For the free-slip boundary condition, the variation of $c_1$ (red squares), $c_2$ (green circles), $c_3$ (blue triangles), and $c_4$ (black diamonds) with $\mathrm{Pr}$ for $\mathrm{Ra} =  2 \times 10^7$. The green curve depicts $c_2/\mathrm{Pr} = 4/\mathrm{Pr} + 0.04$, whereas the black curve represents $c_4 = 1300/\mathrm{Pr} + 150.$}
\label{fig:c_pr_sf}
\end{figure}

After the computation of each of the terms of the momentum equation, we compute the coefficients $c_i$'s that have been defined in Sec.~\ref{sec:Pe}. The $c_i$'s have been plotted in Fig.~\ref{fig:c_ra_sf} as function of $\mathrm{Ra}$, and in Fig.~\ref{fig:c_pr_sf} as function of $\mathrm{Pr}$, and their functional form is tabulated in Table~\ref{table:c_sf}.   The $c_i$'s for the free-slip boundary condition differ in certain ways from those for the no-slip boundary condition.  For the viscous regime (here large $\mathrm{Pr}$) of free-slip flows, $-\nabla \sigma$ is significant.  For the consistency of Eq.~(\ref{eq:Pe_eqn}) we require that $c_1 =0$ and $c_2 \propto \mathrm{Pr}$ in order to cancel $\mathrm{Pr}$ in the $\mathrm{Pr} \rightarrow \infty$ regime.  This is the reason we write $c_2 = -c_2' \mathrm{Pr}$ under the free-slip boundary condition.  For very large $\mathrm{Pr}$, the linear term of $c_2$ dominates its constant counterpart.  Note that for the no-slip boundary condition in the viscous limit, $-\nabla \sigma \approx 0$, and the viscous force and the buoyancy cancel each other.  Hence, the no-slip and the free-slip boundary conditions yield different results.

Let us revisit Eq.~(\ref{eq:Pe_eqn}).  For the viscous regime of the no-slip boundary condition, the nonlinear term and the pressure gradient were negligible, hence we obtained $\mathrm{Pe} \approx (c_3/c_4) \mathrm{Ra}$.  For the free-slip boundary condition, under the viscous regime, $c_2 = -c_2' \mathrm{Pr}$, where $c_2'$ is a positive constant. The sign of $c_2$ is negative because the pressure gradient is along $-\hat{\bf z}$.  Hence
\begin{equation}
c_2' \mathrm{Pe}^2 + c_4 \mathrm{Pe}  -  c_3 \mathrm{Ra} = 0,
\end{equation}
which yields
\begin{equation}
\mathrm{Pe} = \frac{-c_4  + \sqrt{c_4^2  + 4 c_2' c_3\mathrm{Ra}}}{2 c_2'}. \label{eq:Pe_visc_free_slip}
\end{equation}
 Note that the above $\mathrm{Pe}$ is independent of $\mathrm{Pr}$ as observed in numerical simulations~\cite{Pandey:PRE2014}. In the above derivation, $c_2  \propto \mathrm{Pr}$ is an important ingredient.

\begin{table}
\caption{Rayleigh number dependence of various terms of the momentum equation (scaled as $\kappa^2/d^3$) for the free-slip boundary condition. The errors in the exponents are approximately $0.02$.}
\begin{ruledtabular}
\begin{tabular}{ccc}
	& Turbulent regime & Viscous regime \\
\hline 
$|{\bf u} \cdot \nabla {\bf u}|$ & $\mathrm{Ra}^{1.1}$ & $\mathrm{Ra}^{1.3}$ \\
$|(-\nabla \sigma)_{\mathrm{res}}|$ & $\mathrm{Ra}^{1.0}$ & $\mathrm{Ra}^{0.86}$ \\
$|\alpha g \theta_{\mathrm{res}}|$ & $\mathrm{Ra}^{0.89}$ & $\mathrm{Ra}^{0.86}$ \\
$|\nu \nabla^2 {\bf u}|$ & $\mathrm{Ra}^{0.96}$ & $\mathrm{Ra}^{0.89}$ \\
\end{tabular}
\end{ruledtabular}
\label{table:terms_sf}
\end{table}

\begin{table}
\caption{Functional dependence of the coefficients $c_i$'s on $\mathrm{Ra}$ and $\mathrm{Pr}$ for the free-slip boundary condition.}
\begin{ruledtabular}
\begin{tabular}{ccc}
	& $\mathrm{Pr} \leq 9$ & $\mathrm{Pr} > 9$ \\
\hline 
$c_1$ & $0.2 \mathrm{Ra}^{0.20}$ & 5 \\
$c_2$ & $0.05 (4 + 0.04 \mathrm{Pr}) \mathrm{Ra}^{0.15}$ & $22 (6 + 0.28 \mathrm{Pr}) \mathrm{Ra}^{-0.20}$ \\
$c_3$ & $1.35 \mathrm{Ra}^{-0.10} \mathrm{Pr}^{-0.05}$ & 0.30 \\
$c_4$ & $2 \times 10^{-4} (1300/\mathrm{Pr} + 150) \mathrm{Ra}^{0.50}$ & $0.01 (1300/\mathrm{Pr} + 150) \mathrm{Ra}^{0.28}$
\end{tabular}
\end{ruledtabular}
\label{table:c_sf}
\end{table}

\begin{figure}
\includegraphics[scale=1.1]{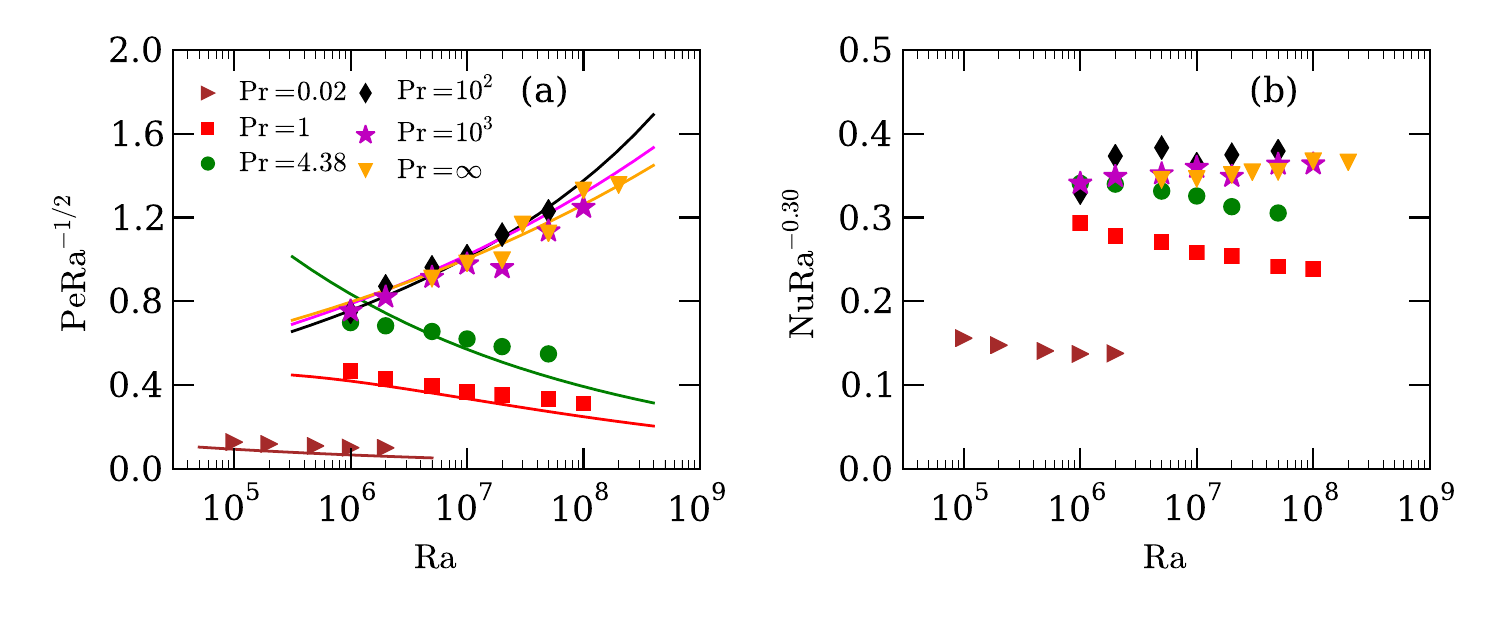}
\caption{For the free-slip boundary condition: (a) the normalized P\'{e}clet number ($\mathrm{PeRa^{-1/2}}$) vs.~$\mathrm{Ra}$. The continuous curves represent analytically computed $\mathrm{Pe}$, which are approximately close to the numerical results.  (b) The normalized Nusselt number ($\mathrm{NuRa^{-0.30}}$) as a function of $\mathrm{Ra}$. For small and moderate $\mathrm{Pr}$, $\mathrm{Pe \sim Ra}^{0.45}$ and $\mathrm{Nu \sim Ra^{0.27}}$, and for very large Prandtl numbers, $\mathrm{Pe \sim Ra}^{0.60}$ and $\mathrm{Nu \sim Ra^{0.32}}$.}
\label{fig:nu_pe_sf}
\end{figure}

In Fig.~\ref{fig:nu_pe_sf}(a) we plot the normalized P\'{e}clet number $\mathrm{PeRa^{-1/2}}$ computed  for various $\mathrm{Pr}$. Here we also plot the analytically computed $\mathrm{Pe}$ [Eq.~(\ref{eq:Pe_analy})] with $c_i$'s from the Table~\ref{table:c_sf} as continuous curves. We observe that our formula fits quite well with the numerical results.  In addition, we also compute $\mathrm{Nu}$, $\theta_{\mathrm{res}}$, $C_{u\theta_{\mathrm{res}}}$, and dissipation rates.  The functional dependence of these quantities with $\mathrm{Ra}$ are listed in Table~\ref{table:scaling_summary_sf}.   Almost all the features are similar to those of the no-slip boundary condition except that $\epsilon_u \propto U^3/d$, similar to unbounded flow, which may be due to weak viscous boundary layer for the free-slip boundary condition. In Fig.~\ref{fig:nu_pe_sf}(b) we plot the normalized Nusselt number computed for the free-slip simulations. As can be observed from the figure, the Nusselt number increases with Prandtl number up to $\mathrm{Pr} = 10^2$ and then it becomes approximately constant. 

In summary, the scaling of large-scale quantities for the no-slip and free-slip boundary conditions have many similarities, but there are certain critical differences.

\begin{table}
\caption{Summary of the scalings for free-slip boundary condition. Quantities are same as those in Table~\ref{table:scaling_summary_ns}.}
\begin{ruledtabular}
\begin{tabular}{ccc}
	& Turbulent regime & Viscous regime \\
\hline 
$C_{u \theta_\mathrm{res}}$ &  $\mathrm{Ra}^{-0.06}$   & $\mathrm{Ra}^{-0.17}$ \\
$\langle \theta_\mathrm{res}^2 \rangle^{1/2} $  & $\mathrm{Ra}^{-0.10}$ &  $\mathrm{Ra}^{-0.12}$\\
$\langle u_z^2 \rangle^{1/2} $  & $\mathrm{Ra}^{0.43}$ &  $\mathrm{Ra}^{0.61}$ \\ 
$\mathrm{Nu}$  & $\mathrm{Ra}^{0.27}$ &  $\mathrm{Ra}^{0.32}$ \\ 
$\epsilon_u$ & $U^3/d$ &  $(\nu U^2/d^2)   \mathrm{Ra}^{0.10}$ \\
$\epsilon_T$ & $(U \Delta^2/d) \mathrm{Ra}^{-0.15}$ &  $(U \Delta^2/d)  \mathrm{Ra}^{-0.29}$ \\
\end{tabular}
\end{ruledtabular}
\label{table:scaling_summary_sf}
\end{table}

\section{Conclusions} \label{sec:concl}

In this paper we derive a general formula for the P\'{e}clet number from the momentum equation. The general formula involves four coefficients that are determined using the numerical data. The predictions from our formula match with most of the past experimental and numerical results.  Our derivation is very different from that of Grossmann and Lohse~\cite{Grossmann:JFM2000, Grossmann:PRL2001, Grossmann:PRE2002} who use exact relations of Shraiman and Siggia~\cite{Shraiman:PRA1990}.  Also, GL's formalism provides 8 different formulae for various limiting cases, but we provide a single formula, whose coefficients are determined using numerical data.

In our paper we also find several other interesting results, which are listed below:
\begin{enumerate}
\item In RBC, the planar average of temperature drops sharply near the boundary layers, and it remains approximately a constant in the bulk.  A consequence of the above observation is that the Fourier transform of the average temperature  $\theta_m$ exhibits $\hat{\theta}_m(0,0,k_z) = -1/(\pi k_z)$, hence the entropy spectrum has a prominent branch $E_\theta(k) \sim k^{-2}$.  The above spectrum has been reported earlier by Mishra and Verma~\cite{Mishra:PRE2010} and Pandey \textit{et al.}~\cite{Pandey:PRE2014}.

\item The modes $\hat{\theta}_m(0,0,k_z)$ do not couple with the velocity modes in the momentum equation. Instead, the momentum equation involves $\theta_\mathrm{res} = \theta - \theta_m$.  It has an important consequence on the scaling of the P\'{e}clet and Nusselt numbers.

\item  The Nusselt number $\mathrm{Nu} = 1+ C_{u\theta_\mathrm{res}} \langle u_z^{2} \rangle_V^{1/2} \langle \theta_\mathrm{res}^{2} \rangle_V^{1/2}$.  The $\mathrm{Ra}$ dependence of  $C_{u\theta_\mathrm{res}}$, $u_z$, and $ \theta_\mathrm{res}$ yields corrections from the ultimate regime scaling $\mathrm{Nu}  \sim \mathrm{Ra}^{1/2}$ to the experimentally-realized  behavior $\mathrm{Nu}  \sim \mathrm{Ra}^{0.3}$.

\item For the no-slip boundary condition we observe that 
\begin{equation}
\frac{\mathrm{Nonlinear~term}}{\mathrm{Viscous~term}} = \frac{|{\bf u \cdot \nabla u}|}{|\nu \nabla^2 {\bf u}|} =  \frac{Ud}{\nu} \frac{c_1}{c_4} \sim \mathrm{Re} \mathrm{Ra}^{-0.14},
\end{equation}
where $c_1 \sim \mathrm{Ra}^{0.10}$ and $c_4 \sim \mathrm{Ra}^{0.24}$.  Thus in RBC, the nonlinear term is weaker than that in free turbulence.  This is due to the wall effect.  The numerical data also reveals that in the turbulent regime, the viscous dissipation rate or the Kolmogorov energy flux $\epsilon_u \sim (U^3/d) \mathrm{Ra}^{-0.21}$, consistent with the suppression of nonlinearity in RBC.  Similarly, the thermal dissipation rate, $\epsilon_T \sim (U \Delta^2/d) \mathrm{Ra}^{-0.19}$.

\item In the viscous regime of RBC, $\epsilon_u \sim (\nu U^2/d^2) \mathrm{Ra}^{0.17}$, thus the viscous dissipation rate is enhanced compared to unbounded flow.

\item Under the free-slip boundary condition, the behavior remains roughly the same as the no-slip boundary condition.  The three main differences between the free-slip and no-slip boundary conditions are
\begin{enumerate}
\item The pressure gradient plays an important role in the viscous regime under the free-slip boundary condition, unlike the no-slip case.  
\item For the free-slip boundary condition, the horizontal components of the pressure gradient and viscous terms are significant, contrary to the no-slip case.
\item For the free-slip case, $\epsilon_u \sim (U^3/d)$ because of the weaker viscous boundary layer. However for the no-slip case, $\epsilon_u \sim (U^3/d) \mathrm{Ra}^{-0.21}$.
\end{enumerate}

\end{enumerate}

In summary, we present the properties of large-scale quantities in RBC, with a focus on the P\'{e}clet number scaling.  These results are very useful for modeling convection in interiors and atmospheres of the planets and stars, as well as in engineering applications.  

\section*{Acknowledgements}

We thank Abhishek Kumar and Anando G. Chatterjee for discussions and help in simulations. We are grateful to the anonymous referees for important suggestions and comments on our manuscript. The simulations were performed on the HPC system and Chaos cluster of  IIT Kanpur, India, and {\sc Shaheen}-II supercomputer of KAUST, Saudi Arabia. This work was supported by a research grant SERB/F/3279/2013-14 from Science and Engineering Research Board, India.

\end{document}